%% file: GEVpaper.tex
\documentclass{aa}
\usepackage{amsmath}
\usepackage{txfonts}
\usepackage{graphicx}
\usepackage{natbib}
\bibpunct{(}{)}{;}{a}{}{,} 
\defcitealias{Smith2007}{S07}
\defcitealias{Beers2000_MWmass}{B00}
\begin{document}
 \title{The RAVE survey: the Galactic escape speed and the mass of the Milky Way}
 \author{     T.~Piffl\inst{1,2}\thanks{email: \texttt{tilmann.piffl@physics.ox.ac.uk}}
	 \and C.~Scannapieco\inst{1}
	 \and J.~Binney\inst{2}
	 \and M.~Steinmetz\inst{1}
	 \and R.-D.~Scholz\inst{1}
	 \and M.~E.~K.~Williams\inst{1}
	 \and R.~S.~{de~Jong}\inst{1}
	 \and G.~Kordopatis\inst{3}
         \and G.~Matijevi\v c\inst{4}
	 \and O.~Bienaym\'{e}\inst{5}
	 \and J.~{Bland-Hawthorn}\inst{6}
	 \and C.~Boeche\inst{7}
         \and K.~Freeman\inst{8}
	 \and B.~Gibson\inst{9}
	 \and G.~Gilmore\inst{3}
	 \and E.~K.~Grebel\inst{7}
	 \and A.~Helmi\inst{10}
	 \and U.~Munari\inst{11}
	 \and J.~F.~Navarro\inst{12}
	 \and Q.~Parker\inst{13,14,15}
	 \and W.~A.~Reid\inst{13,14}
	 \and G.~Seabroke\inst{16}
	 \and F.~Watson\inst{17}
	 \and R.~F.~G.~Wyse\inst{18}
	 \and T.~Zwitter\inst{19,20}
	 }
 \institute{
         Leibniz-Institut f\"ur Astrophysik Potsdam (AIP), An der Sternwarte 16, 14482 Potsdam, Germany		
    \and Rudolf Peierls Centre for Theoretical Physics, 1 Keble Road, Oxford OX1 3NP, UK				
    \and Institute for Astronomy, University of Cambridge, Madingley Road, Cambridge CB3 0HA, UK		
    \and Dept. of Astronomy and Astrophysics, Villanova University, 800 E Lancaster Ave, Villanova, PA 19085, USA 
    \and Observatoire astronomique de Strasbourg, Universit\'{e} de Strasbourg, CNRS, UMR 7550, 11 rue de l'Universi\'{e}, F-67000 Strasbourg, France 
    \and Sydney Institute for Astronomy, University of Sydney, School of Physics A28, NSW 2088, Australia	
    \and Astronomisches Rechen-Institut, Zentrum f\"ur Astronomie der Universit\"at Heidelberg, M\"onchhofstr.\ 12--14, 69120 Heidelberg, Germany 
    \and Research School of Astronomy and Astrophysics, Australian National University, Cotter Rd., Weston, ACT 2611, Australia 
    \and Jeremiah Horrocks Institute, University of Central Lancashire, Preston, PR1 2HE, UK			
    \and Kapteyn Astronomical Institute, University of Groningen, P.O. Box 800, 9700 AV Groningen, The Netherlands 
    \and National Institute of Astrophysics INAF, Astronomical Observatory of Padova, 36012 Asiago, Italy 	
    \and Senior CIfAR Fellow, University of Victoria, Victoria BC, Canada V8P 5C2				
    \and Department of Physics, Macquarie University, Sydney, NSW 2109, Australian				
    \and Research Centre for Astronomy, Astrophysics and Astrophotonics, Macquarie University, Sydney, NSW 2109 Australia 
    \and Australian Astronomical Observatory, PO Box 296, Epping, NSW 1710, Australia				
    \and Mullard Space Science Laboratory, University College London, Holmbury St Mary, Dorking, RH5 6NT, UK 
    \and Australian Astronomical Observatory, PO Box 915, North Ryde, NSW 1670, Australia			
    \and Department of Physics \& Astronomy, Johns Hopkins University, Baltimore, MD 21218, USA 		
    \and University of Ljubljana, Faculty of Mathematics and Physics, Jadranska 19, Ljubljana, Slovenia	
    \and Center of excellence space-si, Askerceva 12, Ljubljana, Slovenia					
    }
 \date{Received ???; Accepted ???}
 \include{abstract}
 \maketitle
 %
 \section{Introduction}
 In the recent years quite a large number of studies concerning the mass of our Galaxy were published. This parameter is of particular interest, because it provides a test for the current cold dark matter paradigm. There is now convincing evidence \citep[e.g.][]{Smith2007} that the Milky Way (MW) exhibits a similar discrepancy between luminous and dynamical mass estimates as was already found in the 1970's for other galaxies. A robust measurement of this parameter is needed to place the Milky Way in the cosmological framework. Furthermore, a detailed knowledge of the mass and the mass profile of the Galaxy is crucial for understanding and modeling the dynamic evolution of the MW satellite galaxies (e.g. \citet{Kallivayalil2013} for the Magellanic clouds) and the Local Group \citep{Marel2012a, Marel2012b}.\\
 Generally, it can be observed, that mass estimates based on stellar kinematics yield low values $\la 10^{12}$~M$_\sun$ \citep{Smith2007, Xue2008, Kafle2012, Deason2012, Bovy2012b}, while methods exploiting the kinematics of satellite galaxies or statistics of large cosmological dark matter simulations find larger values \citep{Wilkinson1999, Li2008, Boylan-Kolchin2011, Busha2011, Boylan2013}. There are some exceptions, however. For example, \citet{Przybilla2010} find a rather high value of $1.7\times10^{12}$~M$_\sun$ taking into account the star J1539+0239, a hyper-velocity star approaching the MW and \citet{Gnedin2010} find a similar value using Jeans modeling of a stellar population in the outer halo. On the other hand \citet{Vera-Ciro2013} estimate a most likely MW mass of $0.8\times10^{12}$~M$_\sun$ analyzing the Aquarius simulations \citep{Springel2008} in combination with semi-analytic models of galaxy formation. \citet{Watkins2010} report an only slightly higher value based on the line of sight velocities of satellite galaxies (see also \citet{Sales2007}), but when they include proper motion estimates they again find a higher mass of $1.4\times10^{12}$~M$_\sun$. Using a mixture of stars and satellite galaxies \citet{Battaglia2005, Battaglia2006} also favor a low mass below $10^{12}$~M$_\sun$. \citet{McMillan2011} found an intermediate mass of $1.3\times10^{12}$~M$_\sun$ including also constraints from photometric data. A further complication of the matter comes from the definition of the total mass of the Galaxy which is different for different authors and so a direct comparison of the quoted values has to be done with care. Finally, there is an independent strong upper limit for the Milky Way mass coming from Local Group timing arguments that estimate the total mass of the combined mass of the Milky Way and Andromeda to $3.2\pm0.6\times10^{12}$~M$_\sun$ \citep{Marel2012a}.\\
 In this work we attempt to estimate the mass of the MW through measuring the escape speed at several Galactocentric radii. In this we follow up on the studies by \citet{Leonard1990}, \citet{Kochanek1996} and \citet{Smith2007} (\citetalias{Smith2007}, hereafter). The latter work made use of an early version of the Radial Velocity Experiment (RAVE; \citet{RAVE_DR1}), a massive spectroscopic stellar survey that finished its observational phase in April 2013 and the almost complete set of data will soon be publicly available in the fourth data release \citep{RAVE_DR4}. This tremendous data set forms the foundation of our study.\\ 
 The escape speed measures the depth of the potential well of the Milky Way and therefore contains information about the mass distribution exterior to the radius for which it is estimated. It thus constitutes a local measurement connected to the very outskirts of our Galaxy. In the absence of dark matter and a purely Newtonian gravity law we would expect a local escape speed of $\sqrt{2}V_\mathrm{LSR} = 311$~km\,s$^{-1}$, assuming the local standard of rest, $V_\mathrm{LSR}$ to be 220~km\,s$^{-1}$ and neglecting the small fraction of visible mass outside the solar circle \citep{Fich1991}. However, the estimates in the literature are much larger than this value, starting with a minimum value of 400~km\,s$^{-1}$ \citep{Alexander1982} to the currently most precise measurement by \citetalias{Smith2007} who find [498,608]~km\,s$^{-1}$ as 90\% confidence range.\\ 
 The paper is structured as follows: in Section~\ref{sec:analysis_technique} we introduce the basic principles of our analysis. Then we go on (Section~\ref{sec:k_from_Aquila}) to describe how we use cosmological simulations to obtain a prior for our maximum likelihood analysis and thereby calibrate our method. After presenting our data and the selection process in Section~\ref{sec:data} we obtain estimates on the Galactic escape speed in Section~\ref{sec:results}. The results are extensively discussed in Section~\ref{sec:discussion} and mass estimates for our Galaxy are obtained and compared to previous measurements. Finally, we conclude and summarize in Section~\ref{sec:conclusions}.
 \section{Methodology} \label{sec:analysis_technique}
 The basic analysis strategy applied in this study was initially introduced by \citet{Leonard1990} and later extended by \citetalias{Smith2007}. They assumed that the stellar system could be described by an ergodic distribution function (DF) $f(E)$ that satisfied $f\to0$ as $E\to\Phi$, the local value of the gravitational potential $\Phi(\vec{r})$. Then the density of stars in velocity space will be a function $n(\varv)$ of speed $\varv$ and tend to zero as $\varv \to \varv_\mathrm{esc} = (2\Phi)^{1/2}$. \citet{Leonard1990} proposed that the asymptotic behavior of $n(\varv)$ could be modeled as
 \begin{equation} \label{eq:LT}
  n(\varv)\propto(\varv_\mathrm{esc}-\varv)^k,
 \end{equation}
 for $\varv < \varv_\mathrm{esc}$, where $k$ is a parameter. Hence we should be able to obtain an estimate of $\varv_\mathrm{esc}$ from a local sample of stellar velocities. \citetalias{Smith2007} used a slightly different functional form
 \begin{equation} \label{eq:S07}
  n(\varv)\propto(\varv_\mathrm{esc}^2-\varv^2)^k = (\varv_\mathrm{esc}-\varv)^k(\varv_\mathrm{esc}+\varv)^k,
 \end{equation}
 that can be derived if $f(E) \propto E^k$ is assumed, but, as we will see in Section~\ref{sec:k_from_Aquila}, results from cosmological simulations are better approximated by Eq.~\ref{eq:LT}. 
 \\
 Currently, the most accurate velocity measurements are line-of-sight velocities, $\varv_\mathrm{los}$, obtained from spectroscopy via the Doppler effect. These measurements have typically uncertainties of a few km\,s$^{-1}$, which is an order of magnitude smaller than the typical uncertainties on tangential velocities obtained from proper motions currently available. \citet{Leonard1990} already showed with simulated data, that because of this, estimates from radial velocities alone are as accurate as estimates that use proper motions as well. The measured velocities $\varv_\mathrm{los}$ have to be corrected for the solar motion to enter a Galactocentric rest frame. These corrected velocities we denote with $\varv_\parallel$.\\
 Following \citet{Leonard1990} we can infer the distribution of $\varv_\parallel$ by integrating over all perpendicular directions:
 \begin{eqnarray} \label{eq:LOS-Velocity-DF}
  n_\parallel(\varv_\parallel~|~\vec{r},k) &\propto& \int\mathrm{d}\vec{v}~ n(\vec{v}~|~\vec{r},k)\delta(\varv_\parallel - \vec{v} \cdot \vec{\hat{m}}) \nonumber \\ 
		&\propto& \left(\varv_\mathrm{esc}(\vec{r}) - |\varv_\parallel|\right)^{k+1}
 \end{eqnarray}
 again for $|\varv_\parallel| < \varv_\mathrm{esc}$. Here $\delta$ denotes the Dirac delta function and $\vec{\hat{m}}$ represents a unit vector along the line of sight.\\
 We do not expect that our approximation for the velocity DF is valid over the whole range of velocities, but only at the high velocity tail of the distribution. Hence we impose a lower limit $\varv_\mathrm{min}$ for the stellar velocities. A further important requirement is that the stellar velocities come from a population that is not rotationally supported, because such a population is clearly not described by an ergodic DF. In the case of stars in the Galaxy, this means that we have to select for stars of the Galactic stellar halo component.\\
 We now apply the following approach to the estimation of $\varv_\mathrm{esc}$. We adopt the likelihood function 
 \begin{equation} \label{eq:LikelihoodEstimator}
  L(\varv_\parallel)=\frac{(\varv_\mathrm{esc}-|\varv_\parallel|)^{k+1}}{\int_{\varv_\mathrm{min}}^{\varv_\mathrm{esc}}\mathrm{d}\varv\,(\varv_\mathrm{esc}-|\varv_\parallel|)^{k+1}} = (k+2)\frac{(\varv_\mathrm{esc}-|\varv_\parallel|)^{k+1}}{(\varv_\mathrm{esc}-\varv_\mathrm{min})^{k+2}}
 \end{equation}
 and determine the likelihood of our catalog of stars that have $|\varv_\parallel|>\varv_\mathrm{min}$ for various choices of $\varv_\mathrm{esc}$ and $k$, then we marginalize the likelihood over the nuisance parameter $k$ and determine the true value of $\varv_\mathrm{esc}$ as the speed that maximizes the marginalized likelihood.
 \subsection{Non-local modeling} \label{sec:non-local_modeling}
 \begin{table}
   \caption{Structural parameters of the baryonic components of our Galaxy model}
   \label{tab:GalaxyModelParameter}
   \centering
   \begin{tabular}{lrl}
    \hline\hline
    disk\\
    \hline
    scale length $R_\mathrm{d}$ & 4 & kpc \\
    scale height $z_\mathrm{d}$ & 0.3 & kpc \\
    mass $M_\mathrm{d}$ & $5 \times 10^{10}$ & M$_\sun$\\
    \hline\hline  
    bulge and stellar halo\\
    \hline
    scale radius $r_\mathrm{b}$ & 0.6 & kpc \\
    mass $M_\mathrm{b}$ & $1.5 \times 10^{10}$ & M$_\sun$\\
    \hline
   \end{tabular}
 \end{table}
 \citet{Leonard1990} (and in a similar form also \citetalias{Smith2007}) used Eq.~\ref{eq:LOS-Velocity-DF} and the maximum likelihood method to obtain constraints on $\varv_\mathrm{esc}$ and $k$ in the solar neighborhood. This rests on the assumption that the stars of which the velocities are used are confined to a volume that is small compared to the size of the Galaxy and thus that $\varv_\mathrm{esc}$ is approximately constant in this volume.\\
 In this study we go a step further and take into account the individual positions of the stars. We do this in two slightly different ways: (1) one can sort the data into Galactocentric radial distance bins and analyze these independently. (2) Alternatively all velocities in the sample are re-scaled to the escape speed at the Sun's position,
 \begin{equation} \label{eq:RescaleVelocities}
  \varv_{\parallel,i}' = \varv_{\parallel,i} \left(\frac{\varv_\mathrm{esc}(\vec{r}_0)}{\varv_\mathrm{esc}(\vec{r}_i)}\right) = \varv_{\parallel,i} \sqrt{\frac{|\Phi(\vec{r}_0)|}{|\Phi(\vec{r}_i)|}},
 \end{equation}
 where $\vec{r}_0$ is the position vector of the Sun. For the gravitational potential, $\Phi(\vec{r})$, model assumptions have to be made. This approach makes use of the full capabilities of the maximum likelihood method to deal with un-binned data and thereby exploit the full information available.\\  
 We will compare the two approaches using the same mass model: a \citet{Miyamoto1975} disk and a \citet{Hernquist1990} bulge for the baryonic components and for the dark matter halo an original or an adiabatically contracted NFW profile \citep{NFW1996,Mo1998}. As structural parameters of the disk and the bulge we use common values that were also used by \citetalias{Smith2007} and \citet{Xue2008} and are given in Table~\ref{tab:GalaxyModelParameter}. The NFW profile has, apart from its virial mass, the (initial) concentration parameter $c$ as a free parameter. In most cases we fix $c$ by requiring the circular speed at the solar radius, $\varv_\mathrm{circ}(R_0)$, to be equal to the local standard of rest, $V_\mathrm{LSR}$ (after a possible contraction of the halo). As a result our simple model has only one free parameter, namely its virial mass. For our results from the first approach using Galactocentric bins we alternatively apply a prior for $c$ taken from the literature to reduce our dependency on the somewhat uncertain value of the local standard of rest.
 \subsection{General behavior of the method} \label{sec:MonteCarloTests}
 \begin{figure}
  \includegraphics{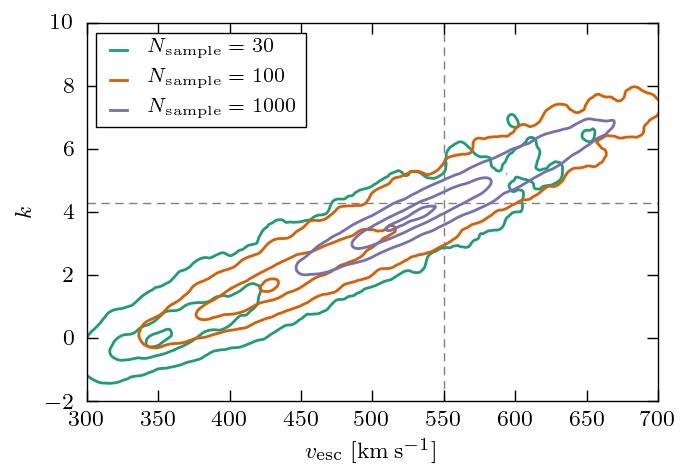}
  \caption{Maximum likelihood parameter pairs computed from mock velocity samples of different sizes. The dotted lines denote the input parameters of the underlying velocity distribution. The contour lines denote positions where the number density fell to 0.9, 0.5 and 0.05 times the maximum value.}
  \label{fig:MonteCarloTests}
 \end{figure}
 To learn more about the general reliability of our analysis strategy we created random velocity samples drawn from a distribution according to Eq.~\ref{eq:LOS-Velocity-DF} with $\varv_\mathrm{esc} = 550$~km\,s$^{-1}$ and $k=4.3$. For each sample we computed the maximum likelihood values for $\varv_\mathrm{esc}$ and $k$. Figure~\ref{fig:MonteCarloTests} shows the resulting parameter distributions for three different sample sizes: 30, 100 and 1000 stars. 5000 samples were created for each value. One immediately recognizes a strong degeneracy between $\varv_\mathrm{esc}$ and $k$ and that the method tends to find parameter pairs with a too low escape speed. This behavior is easy to understand if one considers the asymmetric shape of the velocity distribution. The position of the maximum likelihood pair strongly depends on the highest velocity in the sample -- if the highest velocity is relatively low the method will favor a too low escape speed. This demonstrates the need for additional knowledge about the power-index $k$ as was already noticed by \citetalias{Smith2007}.
 %
 %
 \section{Constraints for \textit{k} from cosmological simulations} \label{sec:k_from_Aquila}
 Almost all of the recent estimates of the Milky Way mass made use of cosmological simulations \citep[e.g.][]{Smith2007, Xue2008, Busha2011, Boylan2013}. In particular, those estimates which rely on stellar kinematics \citep{Smith2007,Xue2008} make use of the realistically complex stellar velocity distributions provided by numerical experiments. In this study we also follow this approach. \citetalias{Smith2007} used simulations to show that the velocity distributions indeed reach all the way up to the escape speed, but more importantly from the simulated stellar kinematics they derived priors on the power-law index $k$. This was fundamental for their study on account of the strong degeneracy between $k$ and the escape speed shown in Figure~\ref{fig:MonteCarloTests} because their data themselves were not enough to break this degeneracy. As we will show later, despite our larger data set we still face the same problem. However, with the advanced numerical simulations available today we can do a much more detailed analysis.\\
 \begin{table}
  \centering
  \caption{Virial radii, masses and velocities after re-scaling the simulations to have a circular speed of 220~km\,s$^{-1}$ at the solar radius $R_0 = 8.28$~kpc.}
  \label{tab:SimScalingTable}
  \begin{tabular}{c c c c c}
   \hline\hline
   Simulation & $R_{340}$ & $M_{340}$ & $V_{340}$ & scaling factor \\
              & (kpc) & ($10^{10}$ M$_\sun$) & (km\,s$^{-1}$) & \\
   \hline
   \input{LT90_figures/AquilaSims/SimScalingTable}
   \hline
  \end{tabular}
 \end{table}
 In this study we make use of the simulations by \citet{Scannapieco2009}. This suite of eight simulations comprises re-simulations of the extensively studied Aquarius halos \citep{Springel2008} including gas particles using a modified version of the Gadget-3 code including star formation, supernova feedback, metal-line cooling and the chemical evolution of the inter-stellar medium. The initial conditions for the eight simulations were randomly selected from a dark matter~only simulation of a much larger volume. The only selection criteria were a final halo mass similar to what is measured for the mass of the Milky Way and no other massive galaxy in the vicinity of the halo at redshift zero. We adopt the naming convention for the simulation runs (A -- H) from \citet{Scannapieco2009}. The initial conditions of simulation~C were also used in the Aquila comparison project \citep{Scannapieco2012}. The galaxies have virial masses between $0.7 - 1.6\times10^{12} M_\sun$ and span a large range of morphologies, from galaxies with a significant disk component (e.g.\ simulations C and G) to pure elliptical galaxies (simulation F). The mass resolution is $0.22$ -- $0.56\times10^6$~M$_\sun$. For a detailed description of the simulations we refer the reader to \citet{Scannapieco2009, Scannapieco2010, Scannapieco2011}. Details regarding the simulation code can be found in \citet{Scannapieco2005, Scannapieco2006} and also in \citet{GadgetPaper}.\\
 An important aspect of the \citet{Scannapieco2009} sample is that the eight simulated galaxies have a broad variety of merger and accretion histories, providing a more or less  representative  sample of Milky Way-mass galaxies formed in a $\Lambda$CDM universe \citep{Scannapieco2011}. Our set of simulations is thus useful for the present study, since it gives us information on the evolution of various galaxies, including all the necessary cosmological processes acting during the formation of galaxies, and at a relatively high resolution.\\
 Also, we note that the same code has been successfully applied to the study of dwarf galaxies \citep{Sawala2011, Sawala2012}, using  the same set of input parameters. Despite a mismatch in the baryon fraction (which is common to almost all simulations of this kind), the resulting galaxies exhibited structures and stellar populations consistent with observations, proving that the code is able to reproduce the formation of galaxies of different masses in a consistent way. Taking into account that the outer stellar halo of massive galaxies form from smaller accreted galaxies, the fact that we do not need to fine-tune the code differently for different masses proves once more the reliability of the simulation code and its results.\\
 To allow a better comparison to the Milky Way we re-scale the simulations to have a circular speed at the solar radius, $R_0=8.28$~kpc, of 220~km\,s$^{-1}$ by the following transformation:
 \begin{equation} \label{eq:scaling}
  \begin{array}{rcl rcl}
  \vec{r}_i' &=& \vec{r}_i / f, &   \vec{v}_i' &=& \vec{v}_i / f, \\
  m_i' &=& m_i / f^3,           &   \Phi_i' &=& \Phi_i / f^2
  \end{array}
 \end{equation}
 with $m_i$ and $\Phi_i$ are the mass and the gravitational potential energy of the $i$th star particle in the simulations. The resulting masses, $M_{340}$, radii, $R_{340}$, and velocities, $V_{340}$ as well as the scaling factors are given in Table~\ref{tab:SimScalingTable}. Throughout this work we use a Hubble constant $H=73$~km\,s$^{-1}$\,Mpc$^{-1}$ and define the virial radius to contain a mean matter density $340\,\rho_\mathrm{crit}$, where $\rho_\mathrm{crit} = 3H^2/8\pi G$ is the critical energy density for a closed universe. The above transformations do not alter the simulation results as they preserve the numerical value of the gravitational constant $G$ governing the stellar motions and also the mass density field $\rho(\vec{r})$ that governs the gas motions as well as the numerical star formation recipe. Only the supernova feedback recipe is not scaling in the same way, but since our scaling factors $f$ are close to unity this is not a major concern.\\
 Since the galaxies in the simulations are not isolated systems, we have to define a limiting distance above which we consider a particle to have escaped its host system. We set this distance to 3$R_{340}$ and set the potential to zero at this radius which results in distances between 430 and 530~kpc in the simulations. This choice is an educated guess and our results are not sensitive to small changes, because the gravitational potential changes only weakly with radius at these distances and in addition, the resulting escape speed is only proportional to the square root of the potential. However, we must not choose a too small value, because otherwise we underestimate the escape speed encoded in the stellar velocity field. On the other hand, we must cut in a regime where the potential is yet not dominated by neighboring (clusters of) galaxies. Our choice is in addition close to half of the distance of the Milky Way and its nearest massive neighbor, the Andromeda galaxy. We further test our choice below. With this definition of the cut-off radius we obtain local escape speeds at $R_0$ from the center between 475 and 550~km\,s$^{-1}$.\\
 Now we select a population of star particles belonging to the stellar halo component. In many numerical studies the separation of the particles into disk and bulge/halo populations is done using a circularity parameter which is defined as the ratio between the particle's angular momentum in the $z$-direction\footnote{The coordinate system is defined such that the disk rotates in the $x-y$-plane.} and the angular momentum of a circular orbit either at the particle's current position \citep{Scannapieco2009,Scannapieco2011} or at the particle's orbital energy \citep{Abadi2003b}. A threshold value is then defined which divides disk and bulge/halo particles. We opt for the very conservative value of $0$ which means that we only take counter-rotating particles. Practically, this is equivalent to selecting all particles with a positive tangential velocity w.r.t\ the Galactic center. This choice allows us to do exactly the same selection as we will do later with the real observational data for which we have to use a very conservative value because of the larger uncertainties in the proper motion measurements.\\
 For similar reasons we also keep only particles in our sample that have Galactocentric distances between 4 and 12~kpc which reflects the range of values of the stars in the RAVE survey which we will use for this study. This further ensures that we exclude particles belonging to the bulge component.\\
 Finally, we set the distance $R_0$ of the observer from the Galactic center to be 8.28~kpc and choose an azimuthal position $\phi_0$ and compute the line-of-sight velocity $\varv_{\parallel,i}$ for each particle in the sample. We further know the exact potential energy $\Phi_i$ of each particle and therefore their local escape speed $\varv_{\mathrm{esc},i}$ is easily computed.\\
 We do this for 4 different azimuthal positions separated from each other by $90\degr$. The positions were chosen such that the inclination angle w.r.t. a possible bar is $45\degr$. The corresponding samples are analyzed individually and also combined. Note, that these samples are practically statistically independent even though a particle could enter two or more samples. However, because we only consider the line-of-sight component of the velocities, only in the unlikely case that a particle is located exactly on the line-of-sight between two observer positions it would gain an incorrect double weight in the combined statistical analysis.
 \\
 \begin{figure}
  \centering
  \includegraphics[width=0.49\textwidth]{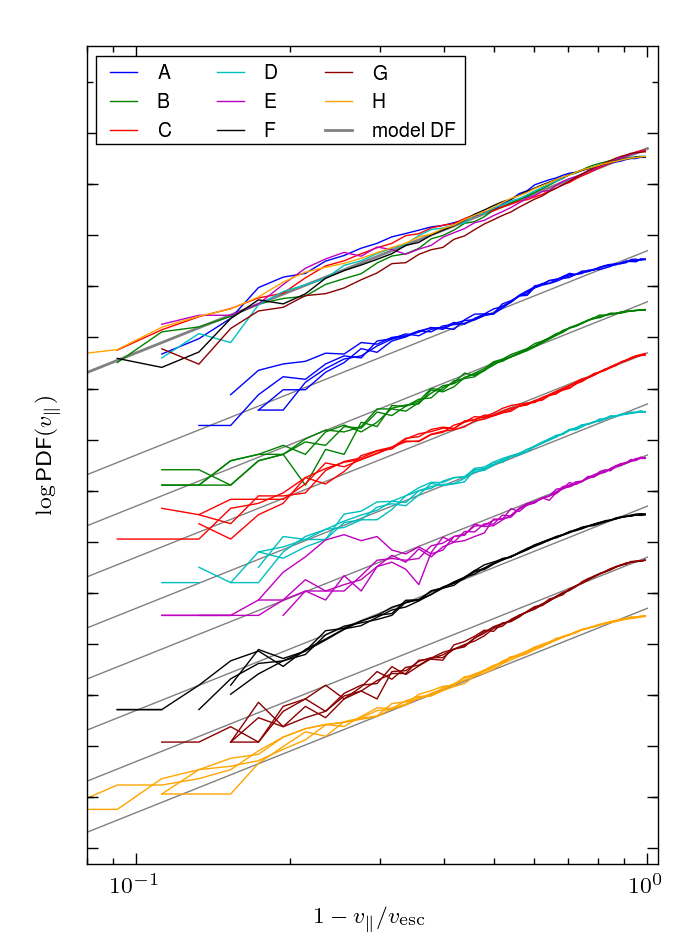}
  \caption{Normalized velocity distributions of the stellar halo population in our eight simulations plotted as a function of $1-\varv_\parallel/\varv_\mathrm{esc}$. Only counter-rotating particles that have Galactocentric distances $r$ between 4 and 12~kpc are considered to select for halo particles (see Section~\ref{sec:Vmin}) and to match the volume observed by the RAVE survey. To allow a comparison each velocity was divided by the escape speed at the particle's position. Different colors indicate different simulations and for each simulation the $\varv_\parallel$ distribution is shown for four different observer positions. The top bundle of curves shows the mean of these four distributions for each simulation plotted on top of each other to allow a comparison. The profiles are shifted vertically in the plot for better visibility. The gray lines illustrate Eq.~\ref{eq:LOS-Velocity-DF} with power-law index $k=3$.}
  \label{fig:LCDM}
 \end{figure}
 \begin{figure}
  \centering
  \includegraphics[width=0.49\textwidth]{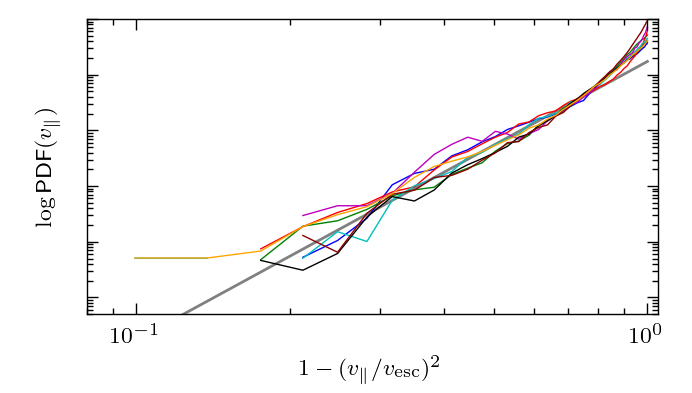}
  \caption{Same as the top bundle of lines in Figure~\ref{fig:LCDM} but plotted as a function of $1-\varv_\parallel^2/\varv_\mathrm{esc}^2$. If the data follows the velocity DF proposed by \citetalias{Smith2007} (gray line) the data should form a straight line in this representation.}
  \label{fig:S07-formalism}
 \end{figure}
 Figure~\ref{fig:LCDM} shows the velocity-space density of star particles as a function of $1-\varv_\parallel/\varv_\mathrm{esc}$ and we see that, remarkably, at the highest speeds these plots have a reasonably straight section, just as \citet{Leonard1990} hypothesized. The slopes of these rectilinear sections scatter around $k=3$ as we will see later.\\
 We also considered the functional form proposed by \citetalias{Smith2007} for the velocity DF, that is $n(\varv) \propto (\varv_\mathrm{esc}^2 - \varv^2)^k$. Figure~\ref{fig:S07-formalism} tests this DF with the simulation data. The curvature implies that this DF does not represent the simulation data as well as the formula proposed by \citet{Leonard1990}.\\
 If we fit Eq.~\ref{eq:LOS-Velocity-DF} to the velocity distributions while fixing $k$ to 3 we recover the escape speeds within 6\%. This confirms our choice of the cut-off radius for the gravitational potential, $3R_{340}$, that was used during the definition of the escape speeds.
 \subsection{The velocity threshold} \label{sec:Vmin}
  \begin{figure}
  \includegraphics{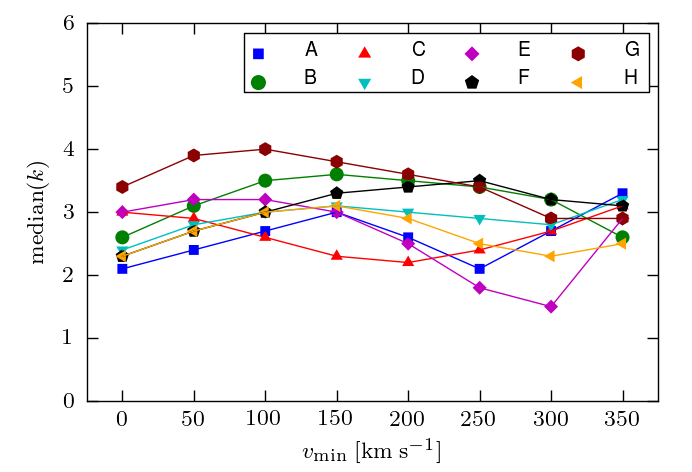}
  \caption{Median values of the likelihood distributions of the power-law index $k$ as a function of the applied threshold velocity $\varv_\mathrm{min}$.}
  \label{fig:Vmin-k-dependency}
 \end{figure}
 We now try to find the best value for the lower threshold velocity $\varv_\mathrm{min}$. \citetalias{Smith2007} had to use a high threshold value for their radial velocities of 300~km\,s$^{-1}$, because the threshold had an additional purpose, namely to select stars from the non-rotating halo component. If one can identify these stars by other means the velocity threshold can be lowered significantly. This adds more stars to the sample and thereby puts our analysis on a broader basis. If the stellar halo had the shape of an isotropic \citet{Plummer1911} sphere, the threshold could be set to zero, because for this model the \citetalias{Smith2007} verison of our approximated velocity distribution function would be exact. However, for other DFs we need to choose a higher value to avoid regions where our approximation breaks down. Again, we use the simulations to select an appropriate value.\\
 We compute the likelihood distribution of $k$ in each simulation using different velocity thresholds using the likelihood estimator
 \begin{equation} \label{eq:L(k)}
  L_\mathrm{tot}(k~|~\varv_\mathrm{min}) = \prod_i L(\varv_{\parallel,i}).
 \end{equation}
 Figure~\ref{fig:Vmin-k-dependency} plots the median values of the likelihood distributions as a function of the threshold velocity. We see a trend of increasing $k$ for $\varv_\mathrm{min}\la 150$~km\,s$^{-1}$ and roughly random behavior above. For low values of $\varv_\mathrm{min}$ simulation G does not follow the general trend. This simulation is the only one in the sample that has a dominating bar in its center \citep{Scannapieco2012_bars} which could contain counter-rotating stars. Given this fact a likely explanation for its peculiar behavior is that with a low velocity threshold, bar particles start entering the sample and thereby alter the velocity distribution.\\
 Simulation E exhibits a dip around $\varv_\mathrm{min} \simeq 300$~km\,s$^{-1}$. A spatially dispersed stellar stream of significant mass is counter-orbiting the galaxy and is entering the sample at one of the observer positions. This is also clearly visible in Figure~\ref{fig:LCDM} as a bump in one of the velocity distributions between 0.2 and 0.3. Furthermore, this galaxy has a rapidly rotating spheroidal component \citep{Scannapieco2009}.\\
 The galaxy in Simulation C has a satellite galaxy very close by. We exclude all star particles in a radius of 3 kpc around the satellite center from our analysis, but there will still be particles entering our samples which originate from this companion and which do not follow the general velocity DF.\\
 All three cases are unlikely to apply for our Milky Way. Our galaxy hosts a much shorter bar and up to now no signatures of a massive stellar stream were found in the RAVE data \citep{Seabroke2008, Williams2011, Antoja2012}. However, it is very interesting to see how our method performs in these rather extreme cases.\\ 
 We adopt threshold velocities $\varv_\mathrm{min} = 200$~km\,s$^{-1}$ and $300$~km\,s$^{-1}$. Both are far enough from the regime where we see systematic evolution in the $k$ values ($\varv_\mathrm{min} \le 150$~km\,s$^{-1}$). For the higher value we can drop the criterion for the particles to be counter-rotating because we can expect the contamination by disk stars to be negligible \citepalias{Smith2007} and thus partly compensate for the reduced sample size.
 \subsubsection{An optimal prior for \textit{k}} \label{sec:prior_for_k}
 %
 %
 From Figure~\ref{fig:Vmin-k-dependency} it seems clear that the different simulated galaxies do not share exactly the same $k$, but cover a considerable range of values. Thus in the analysis of the real data we will have to consider this whole range. We fix the extent of this range by requiring that it delivers optimal results for all four observer positions in all eight simulated galaxies. Hence we applied our analysis to the simulated data by computing the posterior probability distribution
 \begin{equation} \label{eq:PDF_Vesc}
  p(\varv_\mathrm{esc}) \propto \int_{k_\mathrm{min}}^{k_\mathrm{max}}\mathrm{d}k \prod_i L(\varv'_{\parallel,i}~|~\varv_\mathrm{esc},k),
 \end{equation}
 where $L$ was defined in Eq.~\ref{eq:LikelihoodEstimator} and $\varv'_{\parallel,i}$ is the $i$th re-scaled line-of-sight velocity as defined in Eq.~\ref{eq:RescaleVelocities}. We define the median of $p(\varv_\mathrm{esc})$, $\tilde{\varv}_\mathrm{esc}$, as the best estimate. For a comparison of the estimates between different simulations we consider the normalized estimate $\hat{\varv}_\mathrm{esc} = \tilde\varv_\mathrm{esc} / \varv_\mathrm{esc,true}$ with $\varv_\mathrm{esc,true}$ being the true local escape speed in the simulation. By varying $k_\mathrm{min}$ and $k_\mathrm{max}$ we identify those values that minimize the scatter in the sample of 32 $\hat\varv_\mathrm{esc}$ values and at the same time leave the median of the sample close to unity. We find very similar intervals for both threshold velocities and adopt the interval
 \begin{equation}
  2.3 < k < 3.7\,.
 \end{equation}
 Reassuringly, this is very close to the lower part of the interval found by \citetalias{Smith2007} (2.7 -- 4.7) using a different set of simulations. The scatter of the $\hat\varv_\mathrm{esc}\rangle$ values is smaller than 3.5\% ($1\sigma)$ for both velocity thresholds. This scatter cannot be completely explained by the statistical uncertainties of the estimates, so there seems to be an additional uncertainty intrinsic to our analysis technique itself. We will try to quantify this in the next section.
 \subsubsection{Realistic tests} \label{sec:realistic_tests}
 \begin{figure}
  \centering
  \includegraphics{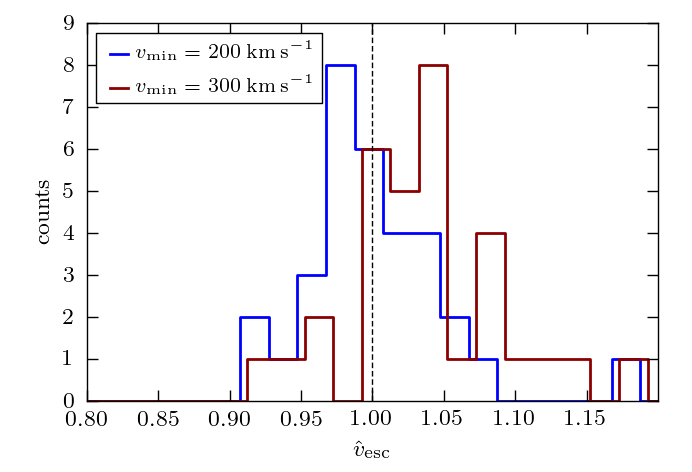}
  \caption{Distribution of $\hat \varv_\mathrm{esc}$ resulting from our 32 test runs of our analysis on simulation data equipped with RAVE-like observational errors and observed in a RAVE-like sky region. In each of the eight simulations four different azimuthal observer positions were tested. A value of unity means an exact recovery of the true local escape speed. The two histograms correspond to our two velocity thresholds applied to the data.}
  \label{fig:RealisticTest}
 \end{figure}
 %
 %
 One important test for our method is whether it still yields correct results if we have imperfect data and a non-isotropic distribution of lines of sights. To simulate typical RAVE measurement errors we attached random Gaussian errors on the parallaxes (distance$^{-1}$), radial velocities and the two proper motion values with standard deviations of 30\%, 3~km\,s$^{-1}$ and 2~mas, respectively. We computed the angular positions of each particle (for a given observer position) and selected only those particles which fell into the approximate survey geometry of the RAVE survey. The latter we define by declination $\delta < 0\degr$ and galactic latitude $|b|>15\degr$.\\
 Figure~\ref{fig:RealisticTest} shows the resulting distributions of $\hat\varv_\mathrm{esc}$ for the two velocity thresholds. Again, the width of the distributions cannot be soly explained by the statistical uncertainties computed from the likelihood distribution, but an additional uncertainty of $\simeq 4\%$ is required to explain the data in a Gaussian approximation. The distribution for $\varv_\mathrm{min}=300$~km\,s$^{-1}$ in addition exhibits a shift to higher values by $\simeq 3\%$. Due to the low number statistics the significance of the shift is unclear ($\sim 3\sigma)$. As we will see in Section~\ref{sec:results}, compared to the statistical uncertainties arising when we analyse the real data it would presents a minor contribution to the overall uncertainty and we neglect the shift for this study.\\
 \begin{figure}
  \centering
  \includegraphics{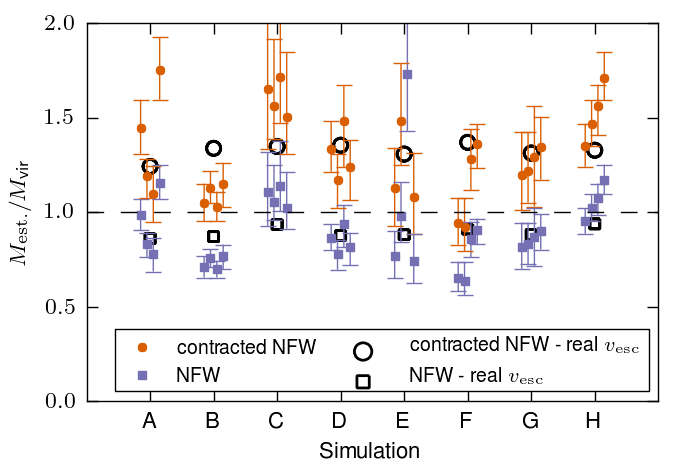}
  \caption{Ratios of the estimated and real virial masses in the eight simulations. For each simulation four mass estimates are plotted based on four azimuthal positions of the Sun in the galaxy. The symbols with error-bars represent the estimates based on the median velocities $\tilde\varv_\mathrm{esc}$ obtained from the error-prone simulation data, while the black symbols show mass estimates for which the real escape speed was used as an input.}
  \label{fig:sim-mass-recovery}
 \end{figure}
 We can go a step further and try to recover the masses of the simulated galaxies using the escape speed estimates. To do this we use the original mass profile of the baryonic components of the galaxies to model our knowledge about the visual parts of the Galaxy and impose an analytic expression for the dark matter halo. As we will do for the real analysis we try two models: an unaltered and an adiabatically contracted NFW sphere. We adjust the halo parameters, the virial mass $M_{340}$ and the concentration $c$, to match both boundary conditions, the circular speed and the escape speed at the solar radius. Figure~\ref{fig:sim-mass-recovery} plots the ratios of the estimated masses and the real virial masses taken from the simulations directly. The adiabatically contracted halo on average over-estimates the virial mass by 25\%, while the pure NFW halo systematically understates the mass by about 15\%. For both halo models we find examples which obtain a very good match with the real mass (e.g. simulation B for the contracted halo and simulation H for the pure NFW halo). However, the cases where the contracted halo yields better results coincide with those cases where the escape speed was underestimated. The colored symbols in Figure~\ref{fig:sim-mass-recovery} mark the mass estimates obtained using the exact escape speed computed from the gravitational potential in the simulation directly. This reveals that the mass estimates from the two halo models effectively bracket the real mass as expected. Note, that we also recover the masses of the three simulations C, E and G that show peculiarities in their velocity distributions. Only for simulation E and one azimuthal observer position do we completely fail to recover the mass. In this case there is a  prominent stellar stream moving along the line of sight.\\
 %
 %
 \section{Data} \label{sec:data}
  \subsection{The RAVE survey}
  The major observational data for this study comes from the fourth data release (DR4) of Radial Velocity Experiment (RAVE), a massive spectroscopic stellar survey conducted using the 6dF multi-object spectrograph on the 1.2-m UK Schmidt Telescope at the Siding Springs Observatory (Australia). A general description of the project can be found in the data release papers: \citet{RAVE_DR1, RAVE_DR2, RAVE_DR3, RAVE_DR4}. The spectra are measured in the \ion{Ca}{II} triplet region with a resolution of $R=7000$. In order to provide an unbiased velocity sample the survey selection function was kept as simple as possible: it is magnitude limited ($9 < I < 12$) and has a weak color-cut of $J-K_s > 0.5$ for stars near the Galactic disk and the Bulge.\\
  In addition to the very precise line-of-sight velocities, $\varv_\mathrm{los}$, several other stellar properties could be derived from the spectra. The astrophysical parameters effective temperature $T_\mathrm{eff}$, surface gravity $\log\,g$ and metallicity [M/H] were multiply estimated using different analysis techniques \citep{RAVE_DR2, RAVE_DR3, Kordopatis2011a}. \citet{Breddels2010}, \citet{Zwitter2010}, \citet{Burnett2011} and \citet{Binney2013} independently used these estimates to derive spectro-photometric distances for a large fraction of the stars in the survey. \citet{Matijevic2012} performed a morphological classification of the spectra and in this way identify binaries and other peculiar stars. Finally \citet{Boeche2011} developed a pipeline to derive individual chemical abundances from the spectra.\\
  The DR4 contains information about nearly 500\,000 spectra of more than 420\,000 individual stars. The target catalog was also cross-matched with other databases to be augmented with additional information like apparent magnitudes and proper motions. For this study we adopted the distances provided by \citet{Binney2013}\footnote{We actually use the parallax estimates, as these are more robust according to \citet{Binney2013}.} and the proper motions from the UCAC4 catalog \citep{UCAC4_paper}.
  \subsection{Sample selection} \label{sec:sample_selection}
  The wealth of information in the RAVE survey presents an ideal foundation for our study. Since \citetalias{Smith2007} the amount of available spectra has grown by a factor of 10 and stellar parameters have become available. The number of high-velocity stars has unfortunately not increased by the same factor, which is most likely due to the fact that RAVE concentrated more on lower Galactic latitudes where the relative abundance of halo stars -- which can have these high velocities -- is much lower.\\
  We use only high-quality observations by selecting only stars which fulfill the following criteria:
  \begin{itemize}
   \item the stars must be classified as 'normal' according to the classification by \citet{Matijevic2012},
   \item the Tonry-Davis correlation coefficient computed by the RAVE pipeline measuring the quality of the spectral fit \citep{RAVE_DR1} must be larger than 10,
   \item the radial velocity correction due to calibration issues \citep[cf.][]{RAVE_DR1} must be smaller than 10~km\,s$^{-1}$,
   \item the signal-to-noise ratio (S/N) must be larger than 25,
   \item the stars must have a distance estimate by \citet{Binney2013},
   \item the star must not be associated with a stellar cluster.
  \end{itemize}
  The first requirement ensures that the star's spectrum can be well fitted with a synthetic spectral library and excludes, among other things, spectral binaries. The last criterion removes in particular the giant star (RAVE-ID J101742.6-462715) from the globular cluster NGC\,3201 that would have otherwise entered our high-velocity samples. Stars in gravitationally self-bound structures like globular clusters, are clearly not covered by our smooth approximation of the velocity distribution of the stellar halo. We further excluded two stars (RAVE-IDs J175802.0-462351 and J142103.5-374549) because of their peculiar location in the Hertzsprung-Russell diagram (green symbols in Figure~\ref{fig:color-metallicity}\footnote{Including these stars does not significantly affect our results.}.\\
  In some cases RAVE observed the same target multiple times. In this case we adopt the measurements with the highest S/N, except for the line-of-sight velocities, $\varv_\mathrm{los}$, where we use the mean value. The median S/N of the high-velocity stars used in the later analysis is 56.\\  
  We then convert the precisely measured $\varv_\mathrm{los}$ into the Galactic rest-frame using the following formula:
  \begin{equation} \label{eq:vGSR}
    \varv_{\parallel,i} = \varv_{\mathrm{los},i} + (U_\sun \cos l_i + (V_\sun+V_\mathrm{LSR}) \sin l_i)\cos b_i + W_\sun \sin b_i,
  \end{equation}
  We define the local standard of rest, $V_\mathrm{LSR}$, to be 220~km\,s$^{-1}$ and for the peculiar motion of the Sun we adopt the values given by \citet{Schoenrich2010}: $U_\sun = 11.1$~km\,s$^{-1}$, $V_\sun = 12.24$~km\,s$^{-1}$ and $W_\sun = 7.25$~km\,s$^{-1}$.\\    
  As mentioned in Section~\ref{sec:analysis_technique} we need to construct a halo sample and we do this in the same way as was done for the simulation data. We compute the Galactocentric tangential velocities, $\varv_\phi$, of all stars in a Galactocentric cylindrical polar coordinate system using the line-of-sight velocities, proper motions, distances and the angular coordinates of the stars. For the distance between the Sun and the Galactic center we use the value $R_0=8.28$ kpc \citep{Gillessen2009b}. We performed a full uncertainty propagation using the Monte-Carlo technique with 2000 re-samplings per star to obtain the uncertainties in $\varv_\phi$. As already done for the simulations we discard all stars with positive median estimate of $\varv_\phi$ and also those for which the upper end of the 95\% confidence interval of $\varv_\phi$ reaches above 100~km\,s$^{-1}$ to obtain a pure stellar halo sample. This is important because a contamination by stars from the rapidly rotating disk component(s) would invalidate our assumptions made in Section~\ref{sec:analysis_technique}. Note, that only for this step we make use of proper motions.\\
  We use the measurements from the UCAC4 catalog \citep{UCAC4_paper} and we avoid entries that are flagged as (projected) double star in UCAC4 itself or in one of the additional source catalogs that are used for the proper motion estimate. In such cases we perform the Monte-Carlo analysis with a flat distribution of proper motions between -50 and 50~mas\,yr$^{-1}$, both in Right Ascension, $\alpha$ and declination, $\delta$.\\  
  In principle, we could also use a metallicity criterion to select halo stars. There are several reasons why we did not opt for this. First, we want to be able to reproduce our selection in the simulations. Unfortunately, the simulated galaxies are all too metal-poor compared to the Milky Way \citep{Tissera2012} and are thus not very reliable in this aspect. This is particularly important in the context of the findings by \citet{Schuster2012} who identified correlations between kinematics and metal abundances in the stellar halo that might be related to different origins of the stars (in-situ formation or accretion). Note, however, that despite the unrealistic metal abundances the formation of the stellar halo is modeled realistically in the simulations including all aspects of accretion and in-situ star formation. In the simulated velocity distributions (Figure~\ref{fig:LCDM}) we do not detect any characteristic features that would indicate that the duality of the stellar halo as found by \citet{Schuster2012} is relevant for our study. Second, we would have to apply a very conservative metallicity threshold in order to avoid contamination by metal-poor disk stars. Because of this our sample size would not significantly increase using a metallicity criterion instead of a kinematic one.\\
  It is worth mentioning, that the star with the highest $\varv_\parallel= -448.8$~km\,s$^{-1}$ in the sample used by \citetalias{Smith2007} (RAVE-ID: J151919.7-191359) did not enter our samples, because it was classified to have problems with the continuum fitting by \citet{Matijevic2012}. \citetalias{Smith2007} showed via re-observations that the velocity measurement is reliable, however, the star did not get a distance estimate from \citet{Binney2013}. \citet{Zwitter2010} estimate a distance of 9.4~kpc which, due to its angular position $(l,b) = (344.6\degr,31.4\degr)$, would place the star behind and above the Galactic center. The star thus clearly violates the assumption by \citetalias{Smith2007} to deal with a locally confined stellar sample and potentially leads to an over-estimate of the escape speed. For the sake of a homogeneous data set we ignored the alternative distance estimate by \citet{Zwitter2010} and discarded the star.\\
  \begin{figure}
   \centering
   \includegraphics[width=0.49\textwidth]{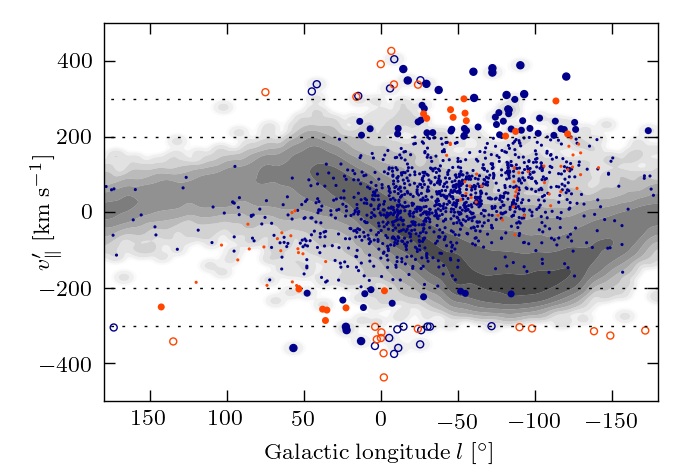}
   \caption{Rescaled radial velocities, $\varv'_\mathrm{r}$, of our high-velocity samples plotted against their Galactic longitudes, $l$. The dashed horizontal lines mark our threshold velocities, $\pm 200$ and $\pm 300$~km\,s$^{-1}$. Blue and orange symbols represent RAVE stars and \citetalias{Beers2000_MWmass} stars, respectively. Open circles mark stars that have $|\varv'_\parallel|>300$~km\,s$^{-1}$, while filled circles represent stars that have $|\varv'_\parallel|>200$~km\,s$^{-1}$ and are classified as halo stars. Colored dots show all stars which we identify as halo stars, i.e. which are on counter-rotating orbits. The gray contours illustrate the complete RAVE mother sample.}
   \label{fig:sample-vGSR-l-plane}
   \centering
   \includegraphics[width=0.49\textwidth]{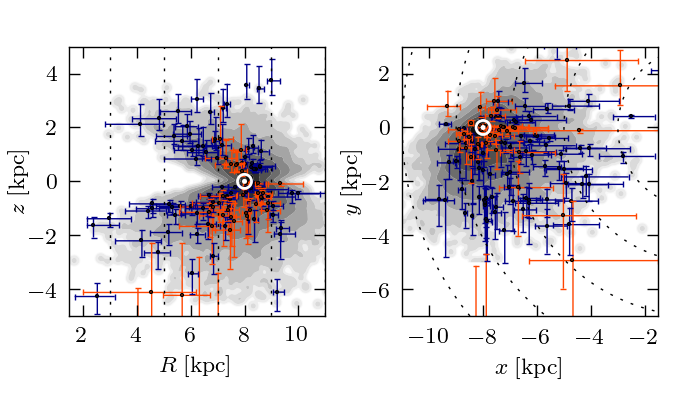}
   \caption{Locations of the stars in our high-velocity sample in the $R$-$z$-plane (left panel) and the $x$-$y$-plane (right panel) as defined in Figure~\ref{fig:sample-vGSR-l-plane}. Blue and orange symbols represent RAVE stars and \citetalias{Beers2000_MWmass} stars, respectively. The error bars show 68\% confidence regions ($\sim1\sigma$). Grey contours show the full RAVE catalog and the position of the Sun is marked by a white '$\sun$'. The dashed lines in both panels mark locations of constant Galactocentric radius $R = \sqrt{x^2+y^2}$.}
   \label{fig:sample-properties}
  \end{figure}
  \begin{figure}
   \centering
   \includegraphics[width=0.49\textwidth]{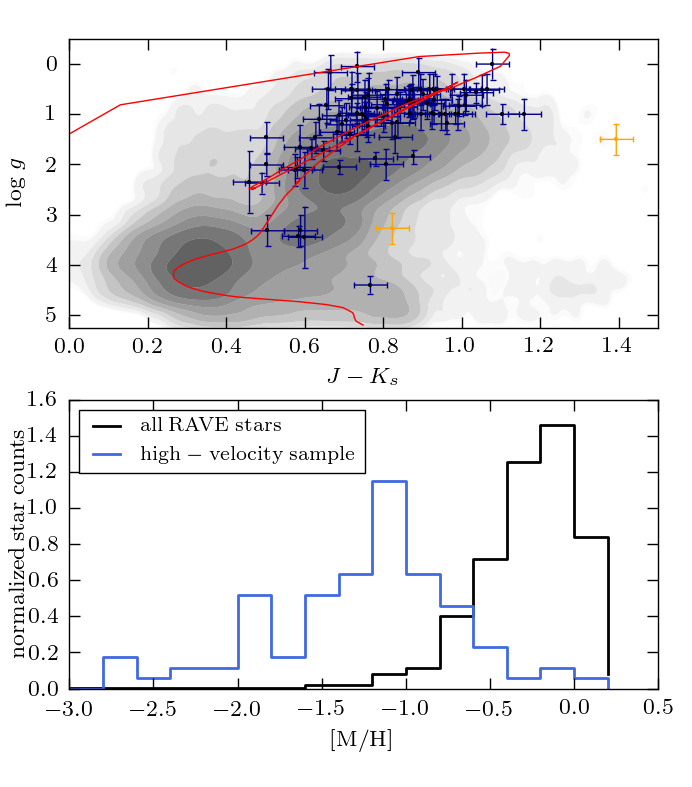}
   \caption{\textit{Upper panel}: Distribution of our high-velocity stars as defined in Figure~\ref{fig:sample-vGSR-l-plane} in a Hertzsprung-Russell diagram (symbols with blue error-bars). For comparison the distribution of all RAVE stars (gray contours) and an isochrone of a stellar population with an age of 10~Gyr and a metallicity of $-1$~dex (red line) is also shown. The two green symbols represent two stars that were excluded from the samples because of the their peculiar locations in this diagram. \textit{Lower panel}: Metallicity distribution of our high-velocity sample (blue histogram). The black histogram shows the metallicity distribution all RAVE stars.}
   \label{fig:color-metallicity}
  \end{figure}
  Figure~\ref{fig:sample-vGSR-l-plane} depicts the velocities $\varv'_\parallel$ of all RAVE stars as a function of Galactic longitude $l$ and the two velocity thresholds $\varv_\mathrm{min} = 200$ and 300~km\,s$^{-1}$. By selecting for a counter-rotating (halo) population (blue dots) we automatically select against the general sinusoidal trend of the RAVE stars in this diagram. Figure~\ref{fig:sample-properties} illustrates the spatial distribution of our high-velocity sample. As a result of RAVE avoiding the low Galactic latitudes, stars with small Galactocentric radii are high above the Galactic plane. Furthermore, because RAVE is a southern hemisphere survey, the stars in the catalog are not symmetrically distributed around the Sun. The stars in our high-velocity sample are mostly giant stars with a metallicity distribution centered at $-1.25$~dex as can be seen in Figure~\ref{fig:color-metallicity}.
  \subsection{Including other literature data} \label{sec:including_beers}
  \begin{figure}
   \centering
   \includegraphics{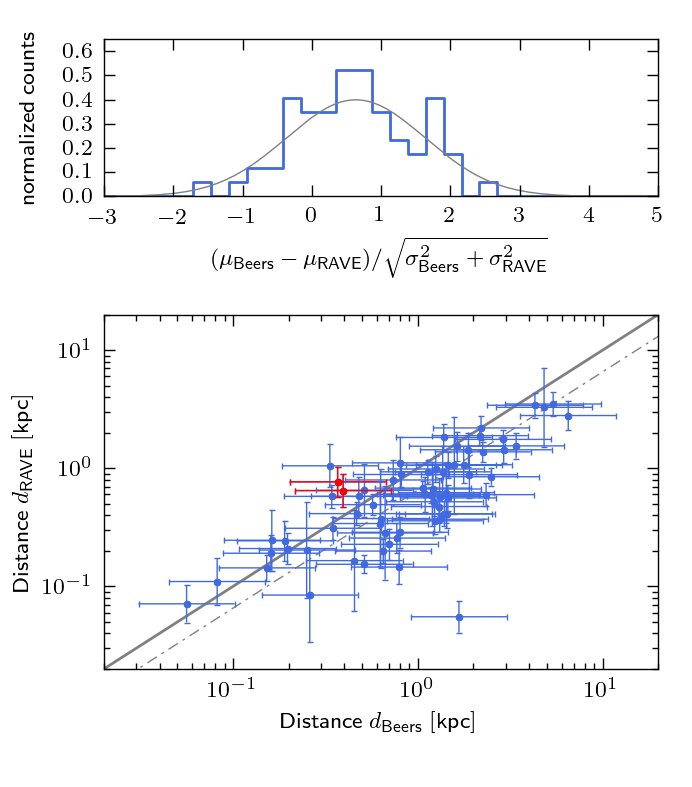}
   \caption{\textit{Upper panel}: Distribution of the differences of the distance modulus estimates, $\mu$, by \citetalias{Beers2000_MWmass} and \citet{Binney2013}, divided by their combined uncertainty for a RAVE-\citetalias{Beers2000_MWmass} overlap sample of 68 stars. With $\sigma_\mathrm{Beers} = 1.3$ mag we find a spread of 1$\sigma$ in the distribution with the median shifted by $0.6\sigma \simeq 0.9$ mag. The grey curve shows a shifted normal distribution. The two red data points mark two stars which were also entering our high-velocity samples. \textit{Lower panel}: Direct comparison of the two distance estimates with $1-\sigma$ error bars. The solid grey line represents equality, while the dashed-dotted line marks equality after reducing the \citetalias{Beers2000_MWmass} distances by a factor of 1.5.}
   \label{fig:RAVE-Beers-Overlap}
  \end{figure}
  To increase our sample sizes we also consider other publicly available and kinematically unbiased data sets. We use the sample of metal-poor dwarf stars collected by \citet[][\citetalias{Beers2000_MWmass} hereafter]{Beers2000_MWmass}. The authors also provide the full 6D phase space information including photometric parallaxes. We updated the proper motions by cross-matching with the UCAC4 catalog \citep{UCAC4_paper}. We found new values for 2011 stars using the closest counterparts within a search radius of 5 arcsec. For ten stars we found two sources in the UCAC4 catalog closer than 5 arcsec and hence discarded these stars. There were further 5 cases where two stars in the \citetalias{Beers2000_MWmass} catalog have the same closest neighbor in the UCAC4 catalog. All these 10  stars were discarded as well. Finally, we kept only those stars with uncertainties in the line-of-sight velocity measurement below 15~km\,s$^{-1}$.\\ 
  There is a small overlap of 123 stars with RAVE, 68 of which have got a parallax estimate, $\varpi$, by \citet{Binney2013} with $\sigma(\varpi) < \varpi$. By chance two of these stars entered our high-velocity samples. This, on the first glance, very unlikely event is not so surprising if we consider our selection for halo stars, the strong bias towards metal-poor halo stars of the \citetalias{Beers2000_MWmass} catalog and the significant completeness of the RAVE survey >50\% in the brighter magnitude bins \citep{RAVE_DR4}.\\
  In order to compare the two distance estimates we convert all distances, $d$, into distance moduli, $\mu = 5\log(d/10~\mathrm{pc})$, because both estimates are based on photometry, so the error distribution should be approximately\footnote{Note, that \citet{Binney2013} actually showed that the RAVE parallax uncertainty distribution is close to normal. However, since both, the RAVE and the \citetalias{Beers2000_MWmass} distances, are based on the apparent magnitudes of the stars, comparing the distance moduli seems to be the better choice, even though the uncertainties are not driven by the uncertainties in the photometry.} symmetric in this quantity. We find that $\sigma_\mathrm{Beers}$ should be about 1.3 mag for the weighted differences (Figure~\ref{fig:RAVE-Beers-Overlap}, upper panel) to have a standard deviation of unity. \citetalias{Beers2000_MWmass} quote an uncertainty of 20\% on their photometric parallax estimates, while our estimate corresponds to roughly 60\%. We adopt our more conservative value and emphasize that this uncertainty is only used during the selection of counter-rotating halo stars.\\
  We further find a systematic shift by a factor $f_\mathrm{dist} = 1.5$ ($\delta\mu = 0.9$ mag) between the two distance estimates, in the sense that the \citetalias{Beers2000_MWmass} distances are greater. Since more information was taken into account to derive the RAVE distances we consider them more reliable. In order to have consistent distances we decrease all \citetalias{Beers2000_MWmass} distances by $f_\mathrm{dist}^{-1}$ and use these calibrated values in our further analysis.\\
  The data set with the currently most accurately estimated 6D phase space coordinates is the Geneva-Copenhagen survey \citep{GCS2004} providing Hipparcos distances and proper motions as well as precise radial velocity measurements. However, this survey is confined to a very small volume around the Sun and is therefore even strongly dominated by disk stars than the RAVE survey. We find only 2 counter-rotating stars in this sample with $|\varv_\parallel|>200$~km\,s$^{-1}$ as well as two (co-rotating) stars with $|\varv_\parallel|>300$~km\,s$^{-1}$. For the sake of homogeneity of our sample we neglect these measurements.
 \section{Results} \label{sec:results}
 \subsection{Comparison to Smith et al. (2007)}
 As a first check we do an exact repetition of the analysis applied by \citetalias{Smith2007} to see whether we get a consistent result. This is interesting because strong deviations could point to possible biases in the data due to, e.g., the slightly increased survey footprint of the sky. RAVE contains 76 stars fulfilling the criteria, which is an increase by a factor 5 (3 if we take the 19 stars from the \citetalias{Beers2000_MWmass}\footnote{Due to the different values of the solar peculiar motion $\vec{U}_\sun$ we have one more star than \citetalias{Smith2007} from this catalog with $|\varv_\parallel|>300$~km\,s$^{-1}$. A further difference is our velocity uncertainty criterion.} catalog into account). The median values of the distributions are effectively the same (537~km\,s$^{-1}$ instead of 544~km\,s$^{-1}$) and the uncertainties resulting from the 90\% confidence interval ([504,574]) are reduced by a factor $0.6$ ($0.7$) for the upper (lower) margin, respectively. If we assume that the precision is proportional to the square root of the sample size we expect a decrease in the uncertainties of a factor $3^{-\frac{1}{2}}~\simeq0.6$.\\
 With the distance estimates available now, we know that this analysis rests on the incorrect assumption that we deal with a local sample. If we apply a distance cut $d_\mathrm{max} = 2.5$~kpc onto the data we obtain a sample of 15 RAVE stars and 16 stars from the \citetalias{Beers2000_MWmass} catalog and we compute a median estimate of $526^{+63}_{-43}$~km\,s$^{-1}$. A lower value is expected because the distance criteria removes mainly stars from the inner Galaxy where stars generally have higher velocities. The reason for this is that RAVE is a southern hemisphere survey and therefore observes mostly the inner Galaxy.
 \subsection{The local escape speed}
 \begin{figure}
  \includegraphics{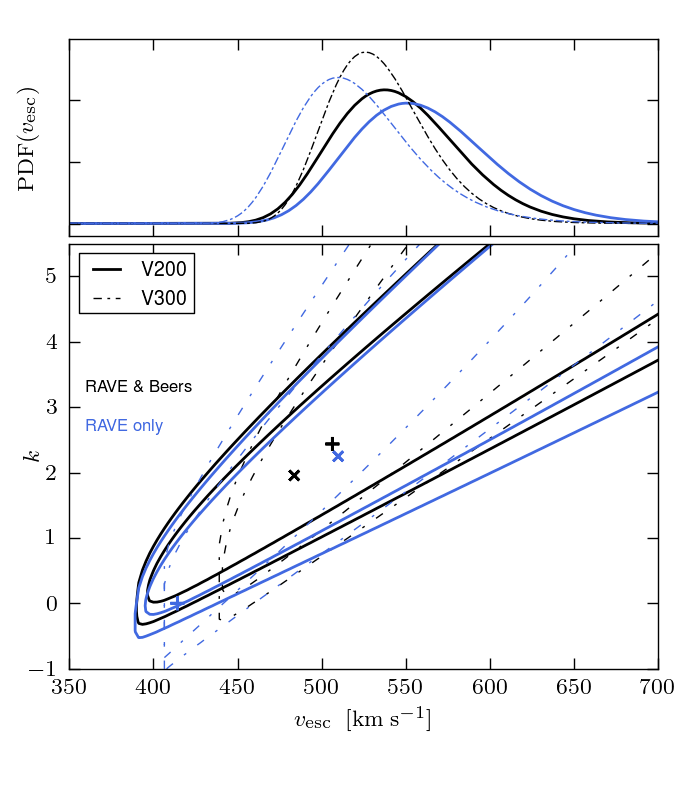}
  \caption{Likelihood distributions of parameter pairs $\varv_\mathrm{esc},k$ (lower panel). The positions of the maximum likelihood pairs are marked with the symbols 'x' for the V200 samples and '+' for the V300 samples. Contour lines mark the locations where the likelihood dropped to 10\% and 1\% of the maximum value. The upper panel shows the likelihood distributions marginalized over the most likely $k$-interval [2.3,3.7]}
  \label{fig:LDF_1D_Localized}
 \end{figure}
 As described as option (2) in Section~\ref{sec:non-local_modeling} we can estimate for all stars in the catalogs what their radial velocity would be if they were situated at the position of the Sun. We then create two samples using the new velocities. For the first sample we select all stars with re-scaled velocities $\varv'_\parallel > 300$~km\,s$^{-1}$. \citetalias{Smith2007} showed that such a high velocity threshold yields predominantly halo stars. The resulting sample contains 53 stars (34 RAVE stars) and we will refer to it as V300. The second sample has a lower velocity threshold of 200~km\,s$^{-1}$, but stars are pre-selected, in analogy to the simulation analysis, considering only stars classified as 'halo' (Section~\ref{sec:sample_selection}). This sample we call V200 and it contains 86 stars (69~RAVE stars). Most of the stars are located closer to the Galactic center than the Sun and thus the correction mostly leads to decreased velocity values. In both samples about 7\% of the stars have repeat observations. The maximum difference between two velocity measurements is 2.5~km\,s$^{-1}$.\\
 The resulting likelihood distribution in the ($\varv_\mathrm{esc},k$) parameter plane is shown in the lower panel of Figure~\ref{fig:LDF_1D_Localized}. The maximum likelihood pairs for the different samples agree very well, except for the pair constructed from RAVE-only V300 sample, which is located near $\varv_\mathrm{esc} \simeq 410$~km\,s$^{-1}$ and $k \simeq 0$. In all cases a clear degeneracy between $k$ and the escape speed is visible. This was already seen by \citetalias{Smith2007} and reflects that a similarly curved form of the velocity DF over the range of radial velocities available by different parameter pairs.\\
 We go further and compute the posterior probability DF for $\varv_\mathrm{esc}$, $p(\varv_\mathrm{esc})$ using Eq.~\ref{eq:PDF_Vesc}, which effectively means that we marginalize over the optimized $k$-interval derived in Section~\ref{sec:prior_for_k}. For the medians of these distributions we obtain higher values than the maximum likelihood value for all samples. This behavior is consistent with our findings in Section~\ref{sec:MonteCarloTests} where we showed that the maximum likelihood analysis tends to yield pairs with too low values of $k$ and $\varv_\mathrm{esc}$. These median values can be found in Table~\ref{tab:results} (''Localized``).
 \subsection{Binning in Galactocentric distance}
 \begin{figure}
  \centering
  \includegraphics{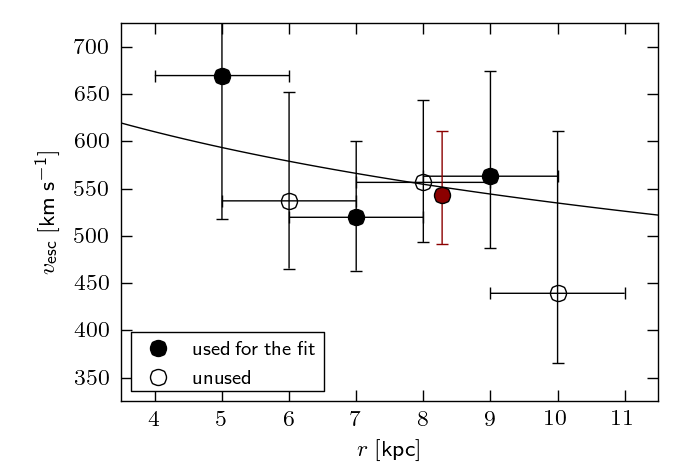}
  \caption{Escape speed estimates and 90\% confidence intervals in Galactocentric radial bins. The solid black line shows our best-fitting model. Only the filled black data points were used in the fitting process. The red data point illustrates the result of our 'localized' approach.}
  \label{fig:BinnedData_r-Vesc}
 \end{figure}
 For halo stars with original $|\varv_\parallel| \ge 200$~km\,s$^{-1}$ we are able to fill several bins in Galactocentric distance $r$ and thereby perform a spatially resolved analysis as described as option (1) in Section~\ref{sec:non-local_modeling}. We chose 6 overlapping bins with a radial width of 2~kpc between 4 and 11~kpc. This bin width is larger than the uncertainties of the projected radius estimates for almost all our sample stars (cf. Figure~\ref{fig:sample-properties}). The number of stars in the bins are 11, 28, 44, 52, 35 and 8, respectively. The resulting median values (again after marginalizing over the optimal $k$-interval) of the posterior PDF and the 90\% confidence intervals are plotted in Figure~\ref{fig:BinnedData_r-Vesc}. The values near the Sun are in very good agreement with the results of the previous section. We find a rather flat escape speed profile except for the out-most bins which contain very few stars, though, and thus have large confidence intervals.
 %
 \section{Discussion} \label{sec:discussion}
 \subsection{Influence of the input parameters}
 The 90\% confidence intervals provided by our analysis technique reflect only the statistical uncertainties resulting from the finite number of stars in our samples. In this section we consider systematic uncertainties. In Section~\ref{sec:realistic_tests} we already showed that our adopted interval for the power-law index $k$ introduces a systematic scatter of about 4\%.\\ 
 A further source of uncertainties comes from the motion of the Sun relative to the Galactic center. While the radial and vertical motion of the Sun is known to very high precision, several authors have come to different conclusions about the tangential motion, $V_\sun$ \citep[e.g.][]{Reid2004, Bovy2012b, Schoenrich2012}. In this study we used the standard value for $V_\mathrm{LSR} = 220$~km\,s$^{-1}$ and the $V_\sun= 12.24$~km\,s$^{-1}$ from \citet{Schoenrich2010}. We repeated the whole analysis using $V_\mathrm{LSR} = 240$~km\,s$^{-1}$ and compared the resulting escape speeds with the values of our standard analysis (cf.\ the lower part of Table~\ref{tab:results}). The magnitudes of the deviations are statistically not significant, but we find systematically lower estimates of the \emph{local} escape speed for the higher value of $V_\mathrm{LSR}$. The shift is close to 20~km\,s$^{-1}$ and thus comparable to the difference $\Delta V_\mathrm{LSR}$. This can be understood if we consider that most stars in the RAVE survey and -- also in our samples -- are observed at negative Galactic longitudes and thus against the direction of Galactic rotation (see Figure~\ref{fig:sample-vGSR-l-plane}). In this case correcting the measured heliocentric line-of-sight velocities with a higher solar tangential motion leads to lower $\varv_\parallel$ which eventually reflects into the escape speed estimate. Note, that this systematic dependency is induced by the half-sky nature of the RAVE survey, while for an all-sky survey this effect might cancel out. In contrast, the exact value of $R_0$ is not influencing our results, as long as it is kept within the range of proposed values around 8~kpc.\\ 
 The quantity with the largest uncertainties used in this study is the heliocentric distance of the stars. In Section~\ref{sec:including_beers} we found a systematic difference between the distances derived for the RAVE stars and for the stars in the \citetalias{Beers2000_MWmass} catalog. Such systematic shifts can arise from various reasons, e.g. different sets of theoretical isochrones, systematic errors in the stellar parameter estimates or different extinction laws. Again we repeated our analysis, this time with all distances increased by a factor 1.5, practically moving to the original distance scale of \citetalias{Beers2000_MWmass}. Again we find a systematic shift to lower local escape speeds of the same order as for alternative value of $V_\mathrm{LSR}$.\\ 
 We finally also tested the influence of the Galaxy model we use to re-scale the stellar velocities according to their spatial position. We changed the disk mass to $6.5\times10^{10}~\mathrm{M}_\sun$ and decreased the disk scale radius to 2.5 kpc, in this way preserving the local surface density of the standard model. The resulting differences in the corrected velocities are below 1\% and no measurable difference in the escape speed estimates were found illustrating the robustness of our methods to reasonable changes in the Galaxy parameters.
 \subsection{A critical view on the input assumptions}
 Our analysis stands and falls with the reliability of our approximation of the velocity DF given in Eq.~\ref{eq:LT}. The conceptual underpinning of this approximation is very weak for four reasons:
 \begin{itemize}
  \item In many analytic equilibrium models of stellar systems at any spatial point there is a non-zero probability density of finding a star right up to the escape speed $\varv_\mathrm{esc}$ at that point, and zero probability at higher speeds. For example the \citet{Jaffe1983} and \citet{Hernquist1990} models have this property but King-Michie models \citep{King1966} do not: in these models the probability density falls to zero at a speed that is smaller than the escape speed. There is hence an important counter-example to the proposition that $n(\varv)$ first vanishes at $\varv=\varv_\mathrm{esc}$.
  \item All theories of galaxy formation, including the standard $\Lambda$CDM paradigm, predict that the velocity distribution becomes radially biased at high speed, so in the context of an equilibrium model there must be significant dependence of the DF on the total angular momentum $J$ in addition to $E$.
  \item As \citet{Spitzer1972} pointed out, in any stellar system, as $E\to0$ the periods of orbits diverge. Consequently the marginally-bound part of phase space cannot be expected to be phase mixed. Specifically, stars that are accelerated to speeds just short of $\varv_\mathrm{esc}$ by fluctuations in $\Phi$ in the inner system take arbitrarily long times to travel to apocenter and return to radii where we may hope to study them. Hence different mechanisms populate the outgoing and incoming parts of phase space at speeds $\varv\sim \varv_\mathrm{esc}$: while parts are populated by cosmic accretion \citep{Abadi2009, Teyssier2009, Piffl2011}, the outgoing part in addition is populated by slingshot processes \citep[e.g.][]{Hills1988, Brown2005} and violent relaxation in the inner galaxy. It follows that we cannot expect the distribution of stars in this portion of phase space to conform to Jeans theorem, even approximately. Yet Eq.~\ref{eq:LT} is founded not just on Jeans theorem but a very special form of it.
  \item Counts of stars in the Sloan Digital Sky Survey (SDSS) have most beautifully demonstrated that the spatial distribution of high-energy stars is very non-smooth. The origin of these fluctuations in stellar density is widely acknowledged to be the impact of cosmic accretion, which ensures that at high energies the DF does not satisfy Jeans theorem.
 \end{itemize}
 From this discussion it should be clear that to obtain a credible relationship between the density of fast stars and $\varv_\mathrm{esc}$ we must engage with the processes that place stars in the marginally bound part of phase space. Fortunately sophisticated simulations of galaxy formation in a cosmological context do just that. Figure~\ref{fig:LCDM} illustrated that Eq.~\ref{eq:LT} catches the general shape of the velocity DF very well. The fact that we find a relatively small interval for the power-law index $k$ that fits all simulated galaxies with their variety of morphologies, argues for the appropriateness of the functional form by \citet{Leonard1990}.\\
 The question remains whether the applied simulation technique influences the range of $k$-values we find, since all eight galaxy models were produced with the same simulation code. In particular, the numerical recipes for so-called sub-grid physics like star formation and stellar energy feedback can have a significant impact on the simulation result as was recently demonstrated in the Aquila code comparison project \citep{Scannapieco2012}. However, the main differences were found in the formation of galaxy disks, while in this study we explicitly focus on the stellar halo that was build up from in-falling satellite galaxies. Differing implementations of sub-grid physics might change the amount of stellar and gas mass being brought in by small galaxies, but it appears unlikely that the phase-space structure of Galactic halo will change significantly. This view is confirmed by the very similar $k$-interval found by \citetalias{Smith2007} using simulations with a completely different implementation of sub-grid physics.
 \subsection{Estimating the mass of the Milky Way} \label{sec:mass_estimate}
 %
 \renewcommand{\arraystretch}{1.4}
 \begin{table*}
  \centering
  \caption{Median and 90\% confidence limits from different analysis strategies. The masses $M_\mathrm{340,NFW}$ are estimated assuming an NFW profile for the dark matter halo and the masses $M_\mathrm{340,contr}$ are based on an adiabatically contracted NFW profile. The upper part of the table shows the results when $V_\mathrm{LSR}$ is assumed to be 220~km\,s$^{-1}$. In the lower part of the table we show the results if we assume a value of 240~km\,s$^{-1}$ to facilitate a comparison to other estimates based on this alternative value.}
  \label{tab:results}
  \begin{tabular}{l c c c | c c c}
  \hline\hline
  Strategy & \multicolumn{3}{c|}{V200} & \multicolumn{3}{c}{V300} \\
           & $\varv_\mathrm{esc}(R_0)$ & $M_\mathrm{340,NFW}$ & $M_\mathrm{340,contr}$& $\varv_\mathrm{esc}(R_0)$ & $M_\mathrm{340,NFW}$ & $M_\mathrm{340,contr}$\\
           & (km\,s$^{-1}$)             & $(10^{12}~\mathrm{M}_\sun)$  & $(10^{12}~\mathrm{M}_\sun)$ & (km\,s$^{-1}$)             & $(10^{12}~\mathrm{M}_\sun)$ & $(10^{12}~\mathrm{M}_\sun)$ \\
  \hline
  \multicolumn{7}{c}{Estimates considering the RAVE and \citetalias{Beers2000_MWmass} data; $V_\mathrm{LSR} = 220$~km\,s$^{-1}$.} \\
  \hline
  \input{ResultsTable_DR4_LT90}
  \hline
  \multicolumn{7}{c}{Estimates considering the RAVE data only; $V_\mathrm{LSR} = 220$~km\,s$^{-1}$.} \\
  \hline
  \input{ResultsTable_DR4_RAVE_LT90}
  \hline
  {}\\
  \hline
  \multicolumn{7}{c}{Estimates considering the RAVE and \citetalias{Beers2000_MWmass} data; $V_\mathrm{LSR} = 240$~km\,s$^{-1}$.} \\
  \hline
  \input{ResultsTable_DR4_LT90_vLSR240}
  \hline
  \multicolumn{7}{c}{Estimates considering the RAVE data only; $V_\mathrm{LSR} = 240$~km\,s$^{-1}$.} \\
  \hline
  \input{ResultsTable_DR4_RAVE_LT90_vLSR240}
  \hline\hline
  \end{tabular}
 \end{table*}
 We now attempt to derive the total mass of the Galaxy using our escape speed estimates. Doing this we exploit the fact that the escape speed is a measure of the local depth of the potential well $\Phi(R_0) = -\frac{1}{2}\varv_\mathrm{esc}^2$. A critical point in our methodology is the question whether the velocity distribution reaches up to $\varv_\mathrm{esc}$ or whether it is truncated at some lower value. \citetalias{Smith2007} used their simulations to show that the level of truncation in the stellar component cannot be more than 10\%. However, to test this they first had to define the local escape speed by fixing a limiting radius beyond which a star is considered unbound. The authors state explicitly that the choice of this radius to be $3R_\mathrm{vir}$ is rather arbitrary. More stringent would be to state that the velocity distribution in the simulations point to a limiting radius of $\sim 3R_\mathrm{vir}$ beyond which stars do not fall back onto the galaxy or fall back only with significantly altered orbital energies, e.g.\ as part of an in-falling satellite galaxy.\\
 It is not a conceptual problem to define the escape speed as the high end of the velocity distribution in disregard of the potential profile outside the corresponding limiting radius. Then it is important, however, to use the same limiting radius while deriving the total mass of the system using an analytic profile. This means we have to re-define the escape speed to
 \begin{equation} \label{eq:re-def:vesc}
  \varv_\mathrm{esc}(r~|~R_\mathrm{max}) = \sqrt{2|\Phi(r) - \Phi(R_\mathrm{max})|}.
 \end{equation}
 $R_\mathrm{max} = 3R_{340}$ seems to be an appropriate value (cf.\ Section~\ref{sec:k_from_Aquila}).\\
 This leads to somewhat higher mass estimates. For example, \citetalias{Smith2007} found an escape speed of 544~km\,s$^{-1}$ and derived a halo mass of $0.85\times10^{12}~M_\sun$ for an NFW profile, practically using $R_\mathrm{max} = \infty$. If one consistently applies $R_\mathrm{max} = 3R_\mathrm{vir}$ the resulting halo mass is $1.05\times10^{12}~M_\sun$, an increase by more than 20\%. This is the reason why our mass estimates are higher than those by \citetalias{Smith2007} even though we find a similar escape speed. Note, that these values represent the masses of the dark matter halo alone while in the remainder of this study we mean the total mass of the Galaxy when we refer to the virial mass $M_{340}$. Keeping this in mind it is then straightforward to compute the virial mass corresponding to a certain local escape speed. As already mentioned, we use the simple mass model presented in Section~\ref{sec:analysis_technique}.\\
 In the case of the escape speed profile obtained via the binned data, the procedure becomes slightly more elaborate. We have to compute the escape speeds at the centers of the radial bins $R_i$ and then take the likelihood from the probability distributions PDF$_{R_i}(\varv_\mathrm{esc})$ in each bin. The product of all these likelihoods\footnote{We only use half of the radial bins in order to have statistically independent measurements.} is the general likelihood assigned to the mass of the model, i.e.
 \begin{equation}
  \hat{L}(M_{340}) = \prod_i \mathrm{PDF}_{R_i}(\varv_\mathrm{esc}(R_i~|~M_{340}))
 \end{equation}
 The results of these mass estimates are presented in Table~\ref{tab:results}. As already seen in Figure~\ref{fig:sim-mass-recovery} for the simulations the adiabatically contracted halo model yields always larger results than the unaltered halo.
 \subsection{Fitting the halo concentration parameter} \label{sec:fit_concentration}
 \begin{figure*}
  \centering
  \includegraphics[width=\textwidth]{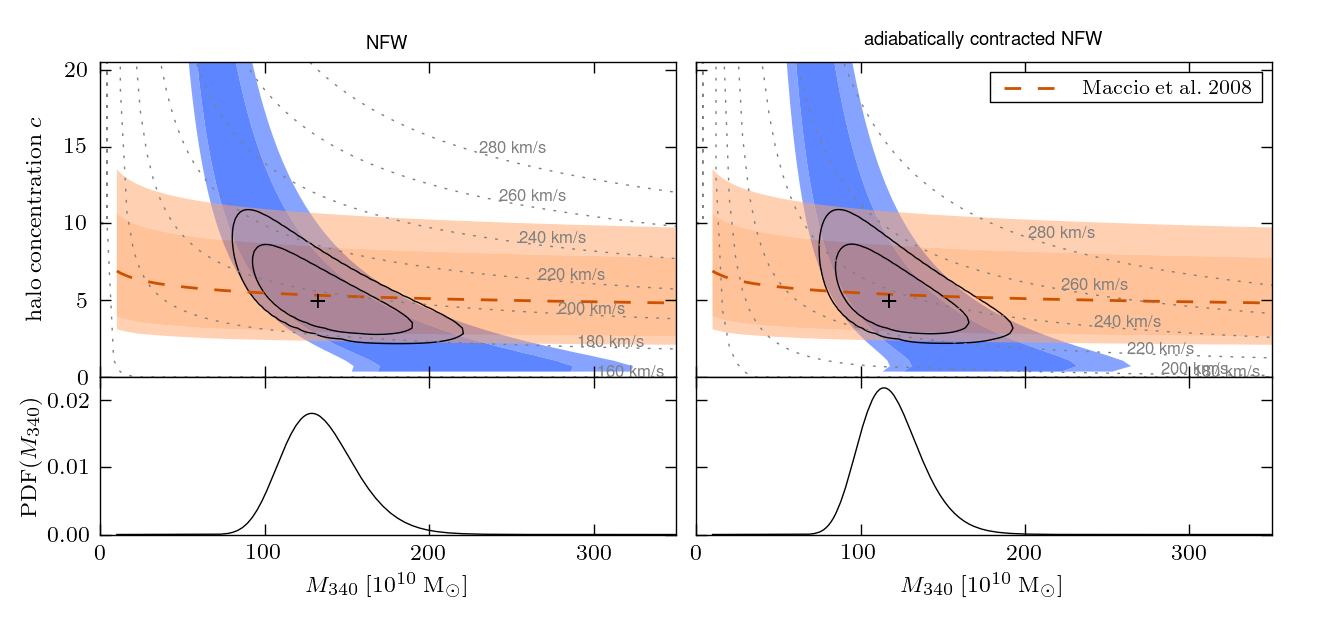}
  \caption{Likelihood distribution resulting from our simple Galaxy model when we leave the halo concentration $c$ (and therefore also $V_\mathrm{LSR}$) as a free parameter (blue area) for an NFW profile as halo model (left panel) and an adiabatically contracted NFW profile (right panel). The red contours arise when we add the constraints on $c$ from cosmological simulations: the relation of the mean $c$ for a given halo mass found by \citet{Maccio2008} is represented by the thick dashed orange line. The orange area illustrates the spread around the mean $c$ values found in the simulations. The different shades in the blue and orange colored areas mark locations where the probability dropped to 10\%, 1\% of the maximum value. Dotted gray lines connect locations with constant circular speed at the solar radius.}
  \label{fig:fitting_mass_and_c}
 \end{figure*}
 Up to now we assumed a fixed value for the local standard of rest, $V_\mathrm{LSR} = 220$~km\,s$^{-1}$, to reduce the number of free parameters in our Galaxy model to one. Recently several authors found larger values for $V_\mathrm{LSR}$ of up to 240~km\,s$^{-1}$ \citep[e.g.][]{Bovy2012b, Schoenrich2012}. If we change the parametrization in the model and use the halo concentration $c$ as a free parameter, we can compute the likelihood distribution in the $(M_{340},c)$-plane in the same way as described in the previous section. Figure~\ref{fig:fitting_mass_and_c} plots the resulting likelihood contours for an NFW halo profile (left panel) and the adiabatically contracted NFW profile (right panel). The solid black curves mark the locations where the likelihood dropped to 10\% and 1\% of the maximum value (which lies near $c\simeq 0$). Grey dotted lines connect locations with common circular velocities at the solar radius.\\
 \citet{NFW1997} showed that the concentration parameter is strongly related to the mass and the formation time of a dark matter halo \citep[see also][]{Neto2007, Maccio2008, Ludlow2012}. With this information we can further constrain the range of likely combinations $(M_{340},c)$. We use the relation for the mean concentration as a function of halo mass proposed by \citet{Maccio2008}. For this we converted their relation for $c_{200}$ to $c_{340}$ to be consistent with our definition of the virial radius. There is significant scatter around this relation reflecting the variety of formation histories of the halos. This scatter is reasonably well fitted by a log-normal distribution with $\sigma_{\log c} = 0.11$ \citep[e.g.][]{Maccio2008, Neto2007}. If we apply this as a prior to our likelihood estimation we obtain the black solid contours plotted in Figure~\ref{fig:fitting_mass_and_c}. Note, that in the adiabatically contracted case the concentration parameters we are quoting are the \emph{initial} concentrations before the contraction. Only these are comparable to results obtained from dark matter-only simulations.\\
 The maximum likelihood pair of values (marked by a black '+' in the figure) for the normal NFW halo is $M_{340} = 1.37\times10^{12}~\mathrm{M}_\sun$ and $c=5$, which implies a circular speed of 196~km\,s$^{-1}$ at the solar radius. The adiabatically contracted NFW profile yields the same $c$ but a somewhat smaller mass of $1.22\times10^{12}~\mathrm{M}_\sun$. Here the resulting circular speed is only 236~km\,s$^{-1}$.\\
 If we marginalize the likelihood distribution along the $c$-axis we obtain the one-dimensional posterior PDF for the virial mass. The median and the 90\% confidence interval we find to be
 \begin{equation}
  M_{340} = 1.3^{+0.4}_{-0.3}\times10^{12}~\mathrm{M}_\sun \nonumber
 \end{equation}
 for the un-altered halo profile. For the adiabatically contracted NFW profile we find
 \begin{equation}
  M_{340} = 1.2^{+0.4}_{-0.3}\times10^{12}~\mathrm{M}_\sun\,, \nonumber
 \end{equation}
 in both cases almost identical to the maximum likelihood value. It is worth noting that in this approach the adiabatically contracted halo model yields the lower mass estimate, while the opposite was the case when we fixed the local standard of rest as done in the previous section.\\
 There are several definitions of the virial radius used in the literature. In this study we used the radius which encompasses a mean density of 340 times the critical density for closure in the universe. If one adopts an over-density of 200 the resulting masses $M_{200}$ increase to $1.6^{+0.5}_{-0.4}\times10^{12}$~M$_\sun$ and $1.4^{+0.4}_{-0.3}\times10^{12}$~M$_\sun$ for the pure and the adiabatically contracted halo profile, respectively. For an over-density of $340\,\Omega_0\sim100$ ($\Omega_0 = 0.3$ being the cosmic mean matter density), as used, e.g., by \citet{Smith2007} or \citet{Xue2008}, the values even increase to $1.9^{+0.6}_{-0.5}\times10^{12}$~M$_\sun$ and $1.6^{+0.5}_{-0.3}\times10^{12}$~M$_\sun$.
 The corresponding virial radii are
 \begin{equation}
  R_{340} = 180\pm20~\text{kpc}\nonumber
 \end{equation}
 for both halo profiles ($R_{200} = 225\pm20$~kpc).
 \subsection{Relation to other mass estimates} \label{sec:other_mass_estimates}
 \begin{figure}
  \centering
  \includegraphics{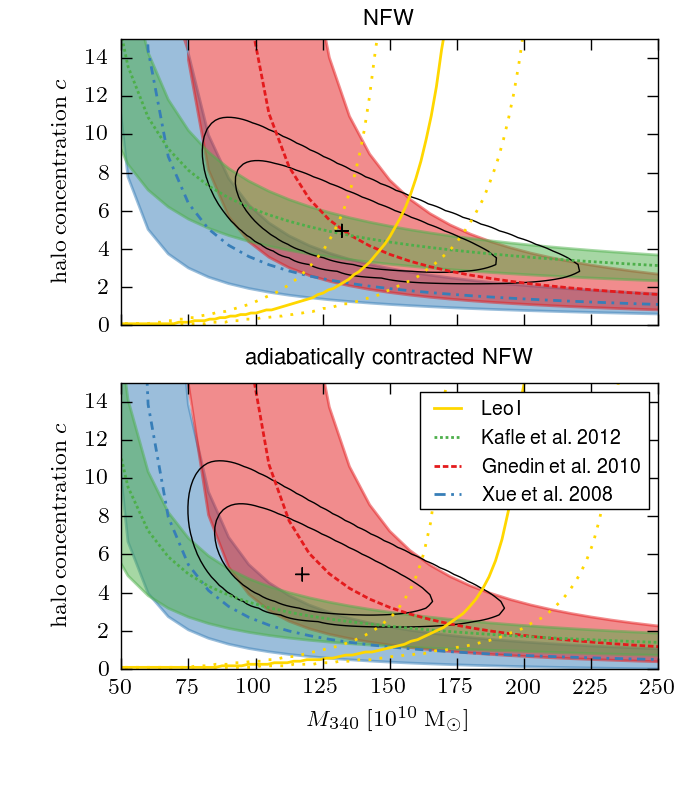}
  \caption{Additional constraints on the parameter pairs ($M_{340},c$) coming from studies from the literature. The black contours are the same as in Figure~\ref{fig:fitting_mass_and_c}. \citet{Gnedin2010} measured the mass interior to 80~kpc from the GC, \citet{Xue2008} interior to 60~kpc and \citet{Kafle2012} interior to 25~kpc. The yellow solid and dotted line separate models for which the satellite galaxy Leo\,I is on a bound orbit (below the lines) from those which it is unbound.}
  \label{fig:Mc-plane-additionalConstraints}
 \end{figure}
 We can include as further constraints literature estimates of total masses interior to various Galactocentric radii by \citet{Xue2008}, \citet{Gnedin2010} and \citet{Kafle2012}. \citet{Gnedin2010} obtained an estimate of a mass of $6.9\times10^{11}$~M$_\sun$ $\pm 20$\% within 80~kpc via Jeans modeling using radial velocity measurements of halo stars between 25 and 80~kpc from the Hypervelocity Star Survey. \citet{Xue2008} found a mass interior to 60~kpc of $4.0\pm0.7\times10^{11}$~M$_\sun$ by re-constructing the circular velocity curve using a radial velocities of halo BHB stars from the SDSS combined with cosmological simulations. \citet{Kafle2012} measured a Galactic mass of $2.1\times10^{11}$~M$_\sun$ interior to 25~kpc from the Galactic center using a similar data set as \citet{Xue2008}, but restricting themselves to stars closer than 25~kpc for which proper motion measurements were available. In this way \citet{Kafle2012} did not have to rely on additional simulation data. We use a 68\% confidence interval of $[1.8,2.3] \times 10^{12}$~M$_\sun$ for this last estimate (green shaded area; P.~Kafle, private communication). Models fulfilling these constraints are marked in Figure~\ref{fig:Mc-plane-additionalConstraints} with colored shaded areas. In the case of the unaltered NFW halo we find an excellent agreement with \citet{Gnedin2010} and \citet{Kafle2012}, while for the adiabatically contracted model the combination of these estimates favor higher virial masses. The estimate by \citet{Xue2008} is only barely consistent with our results on a $1\sigma$-level for both halo models.\\
 Tests with a different model for the Galactic disk ($M_\mathrm{d} = 6.5 \times 10^{10}$~M$_\sun$, $R_\mathrm{d} = 2.5$~kpc, similar to the one used by \citet{Kafle2012} and \citet{Sofue2009}) resulted in decreased mass estimates (10\%), well within the uncertainties. This model changes the values for the circular speed (223~km\,s$^{-1}$ and 264~km\,s$^{-1}$ for the un-altered and the contracted case, respectively) but not the consistency with the mass estimates by \citet{Kafle2012}, \citet{Gnedin2010} or \citet{Xue2008}.\\
 Another important constraint for the Galactic halo is the space motion of the satellite galaxy Leo\,I. \citet{Boylan2013} showed that in the $\Lambda$CDM paradigm it is extremely unlikely that a galaxy like the Milky Way has an unbound close-by satellite galaxy. If we take the recent estimates for the Galactocentric distance of $261\pm13$~kpc and the absolute space velocity of $200^{+22}_{-29}$~km\,s$^{-1}$ \citep{Sohn2013} we can identify those combinations of $M_{340}$ and $c$ that leave Leo\,I on a bound orbit. The line separating models in which Leo\,I is bound from those where it is not bound is also plotted in Figure~\ref{fig:Mc-plane-additionalConstraints}. All models below this line are consistent with a bound orbit of Leo\,I. The dotted lines show the uncertainties in the sense that they mark the ridge lines for the extreme cases that Leo\,I is slower and closer by $1\sigma$ and that it is farther and faster by $1\sigma$. In the case of the un-altered halo profile our mass estimate is consistent with Leo\,I being on a bound orbit, while in the contracted case the mass of the Galaxy would be too low.\\
 Finally, \citet{Przybilla2010} found a star, J1539+0239, with a velocity of $694^{+300}_{-221}$~km\,s$^{-1}$ at a Galactocentric distance of $\sim8$~kpc moving inwards to the Galaxy. The authors argue that this star should therefore be bound to the Milky Way \citep[see also][]{Irrgang2013}. The star is not in the solar vicinity as its heliocentric distance measured to be $12\pm2.3$~kpc, but its Galactocentric distance is comparable to $R_0$. We can therefore directly compare our results. Due to the large uncertainties in the velocity estimate it is not surprising that our most likely value for $\varv_\mathrm{esc}$ is consistent with J1539+0239 to be on a bound orbit. However, if their median velocity is correct, this star is clearly unbound in our model of the Galaxy and must have obtained its high speed via some other mechanism or be of Extragalactic origin.
 \subsection{On the dark matter halo profile}
 The two halo models, un-altered and adiabatically contracted NFW halo, are rather extreme cases and the true shape of the Galactic halo is most likely intermediate to these options \citep{Abadi2010}. When we fixed the circular speed at the Sun's position (as was done for the estimates shown in Table~\ref{tab:results}), the resulting halo masses were strongly dependent on the shape of the profile. However, when we loosened this constraint using a prior on the halo concentration $c$ (as in Section~\ref{sec:fit_concentration}) our mass estimates became fairly robust to changes of the halo model. In this approach the tension between the constraints coming from the circular speed at the solar radius and the mass estimates at larger distances are likely to be resolved by an intermediate halo model as proposed by \citet{Abadi2010}.
 \subsection{Future prospects}
 The ESO cornerstone mission Gaia \citep{Prusti2012} will soon revolutionize the field of Galactic astronomy. It will deliver the full 6D phase space information for more 100 million stars in the extended solar neighborhood. With these data we will not be restricted anymore to the use of radial velocities alone as tangential velocities with similar or even smaller uncertainties will be available. Repeating our analysis with Gaia observations will hence deliver much more precise results.\\
 On the other hand we expect that the full complexity of the Galaxy will appear in these data as well. The comparatively sparse RAVE data allowed to neglect many of the details of the Galactic structure, in particular the clumpy nature of the stellar halo. This might be no longer possible with the Gaia data, or in other words, the precision of the estimate might no longer be limited by the data, but by the assumptions in the analysis method itself. It is hence possible that the gain is smaller than one might expect na\"{i}vely if the analysis is repeated in the exact same manner. More robust knowledge about the structure of the inner galaxy obtained, for example, via the analysis of cold tidal streams \citep{Koposov2010, Sanders2013b} might make it possible to refine these assumptions.
 \section{Conclusions} \label{sec:conclusions}
 In the present study we analyzed the latest data release of the RAVE survey \citep[fourth data release,][]{RAVE_DR4}, together with additional literature data, to estimate the Galactic escape speed ($\varv_\mathrm{esc}$) at various Galactocentric radial bins and through this the virial mass of our Galaxy. For this we define the escape speed as the minimum speed required to reach $3R_{340}$. In order to break a degeneracy between our fitting parameters we had to calibrate our method on a set of cosmological simulations of disk galaxy formation. The 90\% confidence interval for our best estimate of the local escape speed is $492 < \varv_\mathrm{esc} < 587$~km\,s$^{-1}$, with a median value of 533~km\,s$^{-1}$.\\
 With our new $\varv_\mathrm{esc}$ value we can estimate the virial mass of the Galaxy (baryons and dark matter) by assuming a simple mass model of the baryonic content of the Galaxy and a spherical (adiabatically contracted) NFW halo profile and using the local standard of rest, $V_\mathrm{LSR}$, as an additional constraint. The resulting values can be found in Table~\ref{tab:results}.\\ 
 The value of $V_\mathrm{LSR}$ is still under debate. If we loosen our constraint on $V_\mathrm{LSR}$ and and use a prior on the halo concentration parameter, $c$, coming from large cosmological simulations we find a most likely value for the virial mass $M_{340} = 1.3^{+0.4}_{-0.3}\times 10^{12}$~M$_\sun$ for the pure NFW profile and $1.2^{+0.4}_{-0.3}\times 10^{12}$~M$_\sun$ for an adiabatically contracted halo profile.\\
 In Section~\ref{sec:other_mass_estimates} we compare our results to other mass estimates. We find good agreement with estimates based on distant halo stars as well as the space motion of the satellite galaxy Leo\,I.
 %
 \begin{acknowledgements}
 TP and MS acknowledge support by the German Research Foundation under grant DFG-STE-710/4-3. Funding for RAVE has been provided by: the Australian Astronomical Observatory; the Leibniz-Institut f\"ur Astrophysik Potsdam (AIP); the Australian National University; the Australian Research Council; the French National Research Agency; the German Research Foundation (SPP 1177 and SFB 881); the European Research Council (ERC-StG 240271 Galactica); the Istituto Nazionale di Astrofisica at Padova; The Johns Hopkins University; the National Science Foundation of the USA (AST-0908326); the W. M. Keck foundation; the Macquarie University; the Netherlands Research School for Astronomy; the Natural Sciences and Engineering Research Council of Canada; the Slovenian Research Agency; the Swiss National Science Foundation; the Science \& Technology Facilities Council of the UK; Opticon; Strasbourg Observatory; and the Universities of Groningen, Heidelberg and Sydney. The RAVE web site is at \texttt{http://www.rave-survey.org}.\\
 This research made use of the cross-match service provided by CDS, Strasbourg. This research has made use of the SIMBAD database, operated at CDS, Strasbourg, France. This research has made use of the VizieR catalogue access tool, CDS, Strasbourg, France.
 \end{acknowledgements}
 \bibliographystyle{aa}
 \bibliography{all_references}
 
 \appendix
 \section{Defining the potential in the simulations} \label{app:simulation_potential}
 \begin{figure}
  \centering
  \includegraphics[width=0.49\textwidth]{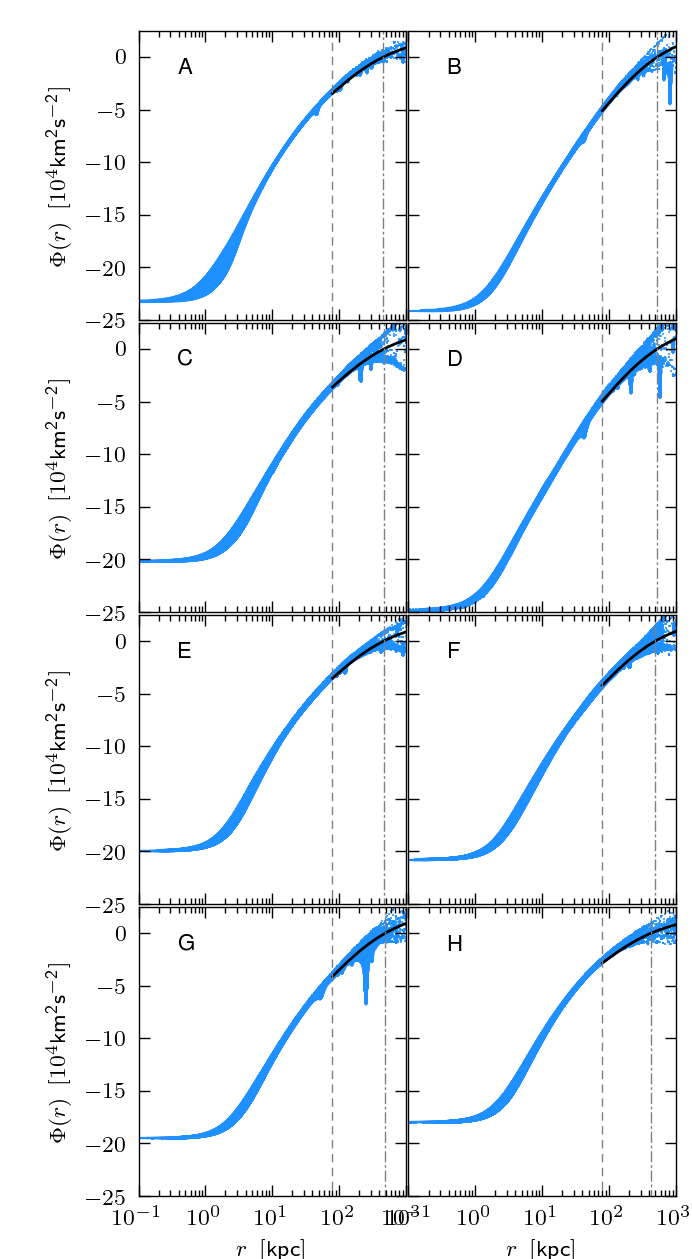}
  \caption{Radial potential profiles of our simulated galaxies. The black line in each panels shows the potential profile of an NFW sphere with the same virial mass as the galaxy and a concentration $c=10$ which was used to define the zero point of the potential.}
  \label{fig:simPotentials}
 \end{figure}
 In this section we briefly describe how we consistently define the potential in each of the 8 simulations we use in this study. Due to the non-spherical symmetry of the mass distribution in the simulation box the gravitational potential shows a spread at a given galactocentric radius. To obtain a robust estimate of the escape speed we redefine the gravitational potential by assuming that the density profile follows a spherically symmetric NFW profile \citep{NFW1997} beyond a radius $r_\mathrm{aux}$:
 \begin{equation} \label{eq:def:potential shift}
  \begin{split}
   \hat{\Phi}(\mathrm{r}) = & \Phi(\mathrm{r}) - \text{median}(\Phi(r_\mathrm{aux})) \\
        &~~~+ \Phi_\mathrm{NFW}(r_\mathrm{aux}) - \Phi_\mathrm{NFW}(3R_{200}) 
  \end{split}
 \end{equation}
 where $\Phi_\mathrm{NFW}$ is the gravitational potential of an NFW sphere with virial mass $10^{10}~\mathrm{M}_\sun$ and concentration $c' = 10$. The radius $r_\mathrm{aux}$ was chosen large enough such that the approximation of an NFW sphere is well justified, but small enough so that the angular variation of the potential is still small. Furthermore there must not be any satellite galaxy near this radius in any of the simulations. For our suite of simulations $r_\mathrm{aux} = 80$ kpc turned out to be a good choice. The value of the concentration parameter is arbitrary as the potential profile is insensitive at large radii for any realistic value of $c'$. Figure~\ref{fig:simPotentials} shows the potential profiles and the approximated profile. The dips in the lower envelopes of the potentials are caused by satellite galaxies orbiting the main halo.\\
 The spread in the potential reflects the fact that these halos live in an anisotropic environment as well as the triaxial shape of the halos. In our simulations we can translate this spread into a maximum deviation from the local escape speed of about 25~km\,s$^{-1}$ which is about 5\%. The maximum is 40~km\,s$^{-1}$ in simulation~D and minimum is 7~km\,s$^{-1}$ for simulation~A.
\end{document}

%% file: abstract.tex
\abstract
{
We construct new estimates on the Galactic escape speed at various Galactocentric radii using the latest data release of the Radial Velocity Experiment (RAVE DR4). Compared to previous studies we have a database larger by a factor of 10 as well as reliable distance estimates for almost all stars. Our analysis is based on the statistical analysis of a rigorously selected sample of 90 high-velocity halo stars from RAVE and a previously published data set. We calibrate and extensively test our method using a suite of cosmological simulations of the formation of Milky Way-sized galaxies. Our best estimate of the local Galactic escape speed, which we define as the minimum speed required to reach three virial radii $R_{340}$, is $533^{+54}_{-41}$~km~s$^{-1}$ (90\% confidence) with an additional 4\% systematic uncertainty, where $R_{340}$ is the Galactocentric radius encompassing a mean overdensity of 340 times the critical density for closure in the Universe. From the escape speed we further derive estimates of the mass of the Galaxy using a simple mass model with two options for the mass profile of the dark matter halo: an unaltered and an adiabatically contracted Navarro, Frenk \& White (NFW) sphere. If we fix the local circular velocity the latter profile yields a significantly higher mass than the uncontracted halo, but if we instead use the statistics on halo concentration parameters in large cosmological simulations as a constraint, we find very similar masses for both models. Our best estimate for $M_{340}$, the mass interior to $R_{340}$ (dark matter and baryons), is $1.3^{+0.4}_{-0.3}\times10^{12}~M_\sun$ (corresponding to $M_{200} = 1.6^{+0.5}_{-0.4}\times10^{12}$~M$_\sun$). This estimate is in good agreement with recently published independent mass estimates based on the kinematics of more distant halo stars and the satellite galaxy Leo\,I.
}

%% file: LT90_figures/AquilaSims/SimScalingTable.tex
  A &             154 &              77 &             147 &   1.20 \\ 
  B &             179 &             120 &             170 &   0.82 \\ 
  C &             157 &              81 &             149 &   1.22 \\ 
  D &             176 &             116 &             168 &   1.05 \\ 
  E &             155 &              79 &             148 &   1.07 \\ 
  F &             166 &              96 &             158 &   0.94 \\ 
  G &             165 &              94 &             157 &   0.88 \\ 
  H &             143 &              62 &             137 &   1.02 \\ 

%% file: ResultsTable_DR4_LT90.tex
              Binned & $557^{+87}_{-63}$ &$1.13^{+0.59}_{-0.35}$ &$1.81^{+1.02}_{-0.62}$ && & \\ 
           Localized & $543^{+67}_{-52}$ &$1.06^{+0.66}_{-0.37}$ &$1.71^{+1.14}_{-0.66}$ &$533^{+54}_{-41}$ &$0.98^{+0.49}_{-0.28}$ &$1.55^{+0.85}_{-0.50}$  \\ 

%% file: ResultsTable_DR4_RAVE_LT90.tex
              Binned & $585^{+109}_{-76}$ &$1.25^{+0.74}_{-0.43}$ &$2.01^{+1.24}_{-0.74}$ && & \\ 
           Localized & $559^{+76}_{-59}$ &$1.19^{+0.82}_{-0.45}$ &$1.94^{+1.41}_{-0.79}$ &$517^{+70}_{-46}$ &$0.86^{+0.60}_{-0.28}$ &$1.35^{+1.05}_{-0.50}$  \\ 

%% file: ResultsTable_DR4_LT90_vLSR240.tex
              Binned & $541^{+93}_{-65}$ &$0.88^{+0.54}_{-0.31}$ &$1.32^{+1.02}_{-0.53}$ && & \\ 
           Localized & $526^{+72}_{-54}$ &$0.76^{+0.53}_{-0.28}$ &$1.09^{+0.97}_{-0.47}$ &$511^{+48}_{-35}$ &$0.67^{+0.30}_{-0.17}$ &$0.94^{+0.54}_{-0.29}$  \\ 

%% file: ResultsTable_DR4_RAVE_LT90_vLSR240.tex
              Binned & $557^{+107}_{-74}$ &$0.95^{+0.68}_{-0.35}$ &$1.47^{+1.25}_{-0.63}$ && & \\ 
           Localized & $535^{+80}_{-57}$ &$0.81^{+0.64}_{-0.31}$ &$1.18^{+1.17}_{-0.52}$ &$483^{+52}_{-37}$ &$0.52^{+0.29}_{-0.15}$ &$0.70^{+0.49}_{-0.24}$  \\ 